\documentclass{aa}  

\usepackage{graphicx}
\usepackage{float}
\usepackage[export]{adjustbox}
\usepackage{txfonts}
\usepackage{xcolor}
\usepackage{xfrac}
\usepackage{fancyhdr}
\usepackage[colorlinks=true,allcolors=blue]{hyperref}%

\newcommand\shorttitle{Turbulent Magnetic Field in the H\,{\sc ii} region Sh 2$-$27}
\newcommand\authors{Raycheva et al.}

\begin{document}

   \title{Turbulent magnetic field in the H\,{\sc ii} region Sh 2$-$27}


   \author{N. C. Raycheva,\inst{1}
          M. Haverkorn,\inst{1}
          S. Ideguchi,\inst{1}
          J. M. Stil,\inst{2}
          B. M. Gaensler,\inst{3}
          X. Sun,\inst{4} 
          J. L. Han,\inst{5,6}
          E. Carretti,\inst{7}\\
          X. Y. Gao,\inst{5}
          T. Wijte\inst{1}
          }

   \institute{Department of Astrophysics/IMAPP, Radboud University, PO Box 9010, 6500 GL, The Netherlands\\ 
              \email{N.Cesur@astro.ru.nl}
        \and     
             Department of Physics and Astronomy, The University of Calgary, 2500 University Drive NW, Calgary, Alberta, T2N1N4, Canada
        \and 
             Dunlap Institute for Astronomy and Astrophysics, 50 St. George Street, Toronto, ON M5S 3H4, Canada
        \and
             School of Physics and Astronomy, Yunnan University, Kunming 650500, China 
        \and    
             CAS Key Laboratory of FAST, NAOC, Chinese Academy of Sciences, Beijing 100101, China
        \and      
             School of Astronomy, University of Chinese Academy of Sciences, Beijing 100049, China
        \and      
             INAF Istituto di Radioastronomia, via Gobetti 101, 40129 Bologna, Italy
             }

   \date{Received 19 September 2020 / Accepted 17 May 2022}

  \abstract
{Magnetic fields in the turbulent interstellar medium (ISM) are a key element in understanding Galactic dynamics, but there are many observational challenges. One useful probe for studying the magnetic field component parallel to the line of sight (LoS) is Faraday rotation of linearly polarized radio synchrotron emission, combined with H$\alpha$ observations. H\,{\sc ii} regions are the perfect laboratories to probe such magnetic fields as they are localized in space, and are well-defined sources often with known distances and measurable electron densities. We chose the H\,{\sc ii} region Sharpless 2$-$27 (Sh 2$-$27) as it is located at intermediate latitudes ($b\sim23\degr$), meaning that it suffers from little LoS confusion from other sources. In addition, it has a large angular diameter ($\sim$10$\degr$), enabling us to study the properties of its magnetic field over a wide range of angular scales.}
{By using a map of the magnetic field strength along the LoS ($B_{\parallel}$) for the first time, we investigate the basic statistical properties of the turbulent magnetic field inside Sh 2$-$27. We study the scaling of the magnetic field fluctuations, compare it to the Kolmogorov scaling, and attempt to find an outer scale of the turbulent magnetic field fluctuations.}
{We used the polarized radio synchrotron emission data from the S-band Polarization All-Sky Survey (S-PASS) at 2.3 GHz, which allowed us to test the impact of Sh 2$-$27 on diffuse Galactic synchrotron polarization. We estimated the rotation measure (\textit{RM}) caused by the H\,{\sc ii} region, using the synchrotron polarization angle. We used the H$\alpha$ data from the Southern H$\alpha$ Sky Survey Atlas to estimate the free electron density ($n_{e}$) in the H\,{\sc ii} region. Using an ellipsoid model for the shape of Sh 2$-$27, and with the observed \textit{RM} and emission measure (\textit{EM}), we estimated the LoS averaged $B_{\parallel}$ for each LoS within the ellipsoid. To characterize the turbulent magnetic field fluctuations, we computed a second-order structure function of $B_{\parallel}$. We compared the structure function to Kolmogorov turbulence, and to simulations of Gaussian random fields processed in the same way as the observations.}
{We present the first continuous map of $B_{\parallel}$ computed using the diffuse polarized radio emission in Sh 2$-$27. We estimate the median value of $n_{e}$ as $7.3\pm0.1$~cm$^{-3}$, and the median value of $B_{\parallel}$ as $-4.5\pm0.1$~$\mu$G, which is comparable to the magnetic field strength in diffuse ISM. The slope of the structure function of the estimated $B_{\parallel}$-map is found to be slightly steeper than Kolmogorov, consistent with our Gaussian-random-field $B_{\parallel}$ simulations revealing that an input Kolmogorov slope in the magnetic field results in a somewhat steeper slope in $B_{\parallel}$. These results suggest that the lower limit to the outer scale of turbulence is 10 pc in the H\,{\sc ii} region, which is comparable to the size of the computation domain.}
{The structure functions of $B_{\parallel}$ fluctuations in Sh 2$-$27 show that the magnetic field fluctuations in this H\,{\sc ii} region are consistent with a Kolmogorov-like turbulence. Comparing the observed and simulated $B_{\parallel}$ structure functions results in the estimation of a lower limit to the outer scale of the turbulent magnetic field fluctuations of 10~pc, which is limited by the size of the field of view under study. This may indicate that the turbulence probed here could actually be cascading from the larger scales in the ambient medium, associated with the interstellar turbulence in the general ISM, which is illuminated by the presence of Sh 2$-$27. }


   \maketitle
%

\section{Introduction}\label{sec:intro}
Understanding Galactic magnetism is crucial as it plays an important role in  the physical properties of the interstellar medium (ISM). Through the electrical conductivity of magnetized astrophysical plasma, magnetic field fluctuations are coupled with turbulence in the ISM. Therefore, studying magnetic field properties will help to shed light on interstellar turbulence, for example  through its outer scale of fluctuations (see reviews in  \citealt{Elmegreen2004,Lazarian2004,McKee2007,Lazarian2009}). For decades it has been known that turbulence is important to many astrophysical processes, such as cosmic ray propagation (e.g. \citealt{Chevalier1984,Minter1996,Giacalone2017}), amplification of magnetic fields (e.g. \citealt{DeYoung1980,Brandenburg2005,Martin-Alvarez2018}), modelling of molecular clouds (e.g. \citealt{Mestel1956,Ostriker2001,King2019}), star formation (e.g. \citealt{Ferriere2001,Li2009,Wurster2018}), and heating of the ISM (e.g. \citealt{Minter1997,Spangler2007,Pan2009}). 

Turbulence in the ISM induces coherent fluctuations on many scales \citep{Scalo1984}, and its non-linear dynamical behaviour indicates that its characteristics cannot be expressed easily, although they can be studied statistically \citep{Miville1995}. To quantify turbulence fluctuations in the ISM through Faraday rotation\footnote{In a magneto-ionic plasma, a travelling electromagnetic wave with an initial position angle ($\chi_{0}$) is rotated by Faraday rotation to a polarization angle ($\chi$) at wavelength $\lambda$ as \(\chi(\lambda^{2})=\chi_{0}+RM\lambda^{2}\), where \textit{RM} stands for rotation measure \citep{Burn1966}, which  is defined as
\begin{equation}
\left(\frac{RM(d)}{\rm rad\,m^{-2}} \right) \ = \ 0.812\intop_{0}^{d}\left(\frac{n_{e}(s)}{\textrm{cm}^{-3}}\right)\left(\frac{{B_{\parallel}(s)}}{\mu\textrm{G}}\right) \left(\frac{ds}{\textrm{pc}}\right),
\label{eq:RM}
\end{equation}
where $d$  is the path length from a polarized source to the observer, $n_{e}$ is the free electron density, $B_{\parallel}$ is the LoS magnetic field strength, and $ds$ is the incremental displacement along a LoS to a source, which is estimated through the thickness of the H\,{\sc ii} region along the LoS. The corresponding sign of \textit{RM} gives the direction of $B_{\parallel}$:  a positive \textit{RM} represents a magnetic field directed to the observer and vice versa.}, structure functions are commonly used (e.g. \citealt{Haverkorn2004-1,Haverkorn2006b,Mao2010,Stil2011}). 
ISM turbulence studies through Faraday rotation are generally difficult to interpret. One of the reasons for this is that the Faraday rotation involves an integration over an unknown path length; it is weighted by electron density, which is calculated from integrated measurements along a path length that is not necessarily the same. One way to study turbulence in a confined region where both path length and local electron density are fairly well defined is to probe an H\,{\sc ii} region. Within the context of this work, local Faraday rotation and electron density measurements are possible by subtracting the  foreground and background. In addition, the path length can be accurately estimated if the distance to the H\,{\sc ii} region is known, as this paper aims to show. Although in principle, this probes turbulent properties inside the H\,{\sc ii} region, \citet{Spangler2021} argues that magnetic fields in H\,{\sc ii} regions  carry information from the general interstellar magnetic field. Thus, although we are only investigating magnetic fields inside an H\,{\sc ii} region, the results may well be significant to the general ISM.

Magnetic fields in H\,{\sc ii} regions have been studied using polarized synchrotron emission and their Faraday rotation. Magnetic field strengths have been derived from depolarization by H\,{\sc ii} regions \citep{Gray1999,Sun2007,Gao2010,Xiao2011} or from measurements of Faraday rotation of extragalactic background sources \citep{Whiting2009,Stil2011,Harvey-Smith2011}. The latter studies show enhanced rotation measure (\textit{RM}) in H\,{\sc ii} regions, but conclude that the calculated magnetic field values are barely larger than interstellar values. This leads to the conclusion that it is primarily an increase in electron density, rather than an enhanced magnetic field, that causes high \textit{RM}s \citep{Costa2016}.

Faraday rotating regions that do not significantly radiate synchrotron emission by themselves but change the angle of the polarized emission passing through them are known as Faraday screens \citep{Haverkorn2003a,Shukurov2003}. In particular, H\,{\sc ii} regions are Faraday screens to background continuum radiation passing through them (e.g. \citealt{Sun2007,Gao2010,Harvey-Smith2011}).

This study aims to contribute to understanding turbulent fluctuations in the magnetic field of an H\,{\sc ii} region by using structure functions of second order on a $B_{\parallel}$-map. We use polarized radio emission from the S-band Polarization All-Sky Survey\footnote{\url{https://sites.google.com/inaf.it/spass}} (S-PASS) at 2.3 GHz \citep{Carretti2019} to determine \textit{RM}s. These are converted into $B_{\parallel}$ using H$\alpha$ density data from the Southern H$\alpha$ Sky Survey Atlas\footnote{\url{http://amundsen.swarthmore.edu/}} (SHASSA, \citealt{Gaustad2001}) as a tracer of electron density, and corrected for dust reddening from extinction maps\footnote{\url{https://irsa.ipac.caltech.edu/applications/DUST/}} \citep{Schlegel1998, Schlafly2011}. This results in the first-ever continuous magnetic field map of an H\,{\sc ii} region, based on diffuse emission. With the structure functions, we endeavour to derive the turbulent slope and outer scale of fluctuations.

The paper proceeds as follows. We give the basic properties of the H\,{\sc ii} region we studied in Sect.~\ref{sec:sh2-27}. We introduce the data used in this study in Sect.~\ref{sec:data}. We present our methods and applications in Sect.~\ref{sec:method}. In Sect.~\ref{sec:analysis_interp} we introduce the radio polarization observations used in this study, which led us to estimate $B_{\parallel}$. We study the statistical characteristics of the turbulence in Sect.~\ref{sec:turbulence} with observed and simulated data. We discuss the results in Sect.~\ref{sec:discussion}, and present
our conclusions in Sect.~\ref{sec:conc}.

\section{Properties of Sh 2--27}\label{sec:sh2-27}
 The Galactic H\,{\sc ii} region Sharpless 2$-$27 (Sh 2$-$27) was discovered by \citet{Sharpless1959}. It is ionized by the runaway star $\zeta$ Oph (\citealt{Blaauw1961}) located at $(l, b)=(6{\fdg}3, +23{\fdg}6)$. The distance from the Sun to Sh 2$-$27 has been estimated as 180 pc \citep{Gaia2016,Gaia2018}. Sh 2$-$27 is a Faraday screen  \citep*{Iacobelli2014,Robitaille2017,Robitaille2018} and a good object to study as it is considerably off the Galactic plane (centred at $b\sim23\degr$); it thus suffers  little from line of sight (LoS) confusion from other sources, and provides an attainable wide range of angular scales, allowing us to investigate its polarization properties. Here we use its Faraday screen property to probe $B_{\parallel}$ in Sh 2$-$27 by estimating a diffuse \textit{RM}. Some dark clouds \citep{Lynds1962} are located in the foreground, which might overlap with the H\,{\sc ii} region \citep{Sivan1974,Tachihara2012,Choi2015}. The CO contours of these dark clouds from \citet{Dame2001} can be seen in Fig.~\ref{fig:regions}, and we explain how we treat them in Sect.~\ref{sec:em}. 
 
 \subsection{Magnetic field properties of Sh 2--27}
 Using Faraday rotation of polarized radio sources in the background from \citet{Taylor2009} and SHASSA data, \citet{Harvey-Smith2011} found a median |$B_{\parallel}$| as $\sim$6.1 $\mu$G in Sh 2$-$27. However, since they only had 57 polarized background sources behind Sh 2$-$27, they constructed neither structure functions nor turbulence properties in this H\,{\sc ii} region. Using the polarization gradient technique on S-PASS data, \citet{Iacobelli2014} suggested the presence of turbulent fluctuations in Sh 2$-$27. They interpreted the high gradients in the linear polarization vector in Sh 2$-$27 as strong turbulent fluctuations or weak shocks, but could not characterize this turbulence any further. \citet{Robitaille2017} also computed the gradient of linearly polarized synchrotron emission with the S-PASS data, and discussed the foreground Faraday fluctuations in Sh 2$-$27, as well as E- and B-mode maps, but did not use \textit{RM}s and could not determine any magnetic field strengths. \citet{Thomson2019} used polarized radio observations at 300$-$480 MHz from the low-band Southern Global Magneto-Ionic Medium Survey \citep{Wolleben2019}. These low frequencies allowed them to estimate the  $B_{\parallel}$ of the H\,{\sc ii} region's foreground and of the Local Bubble because the diffuse polarized emission from the H\,{\sc ii} region itself is completely depolarized at these low frequencies.
 
\section{Data}\label{sec:data}
\subsection{Radio polarization}
This work is based on the data from S-PASS, a highly sensitive single-dish polarimetric survey of the entire southern sky at 2.3 GHz. The main observational details can be found in Table~1 of \citet{Carretti2019}. S-PASS was performed with the Parkes 64 m Radio Telescope \textit{Murriyang} and S-band (13 cm) receiver. Its angular resolution is 8$\farcm$9 and the pixel size is 3$\farcm$4. We used the Stokes $Q$ and $U$ maps produced by this survey to estimate polarization angles, which can lead us to calculate $B_{\parallel}$-map. The de-biased linear polarized radio intensity of Sh 2$-$27 is calculated as $P=(Q^{2}+U^{2}-1.2\sigma^{2}_{QU})^{1/2}$ \citep{Wardle1974,Vaillancourt2006}. The associated noise, $\sigma_{QU}$, was taken from the sensitivity map of S-PASS, assuming that $\sigma_{Q}$ and $\sigma_{U}$ are identical (see Appendix~\ref{app:app}).

\subsection{H\texorpdfstring{$\alpha$}{a} intensity} 
As a probe for the electron density, we used the smoothed continuum-subtracted map of H$\alpha$ surface brightness ($I_{H\alpha}$) data from SHASSA, which has a noise of $\sim$0.7 Rayleigh (1 $R=10^{6}/4\pi$~photons~cm$^{-2}$~s$^{-1}$~sr$^{-1}$) and an angular resolution of 4${\arcmin}$. The map was additionally smoothed to the S-PASS resolution, and resampled onto the S-PASS pixel size using the Python package {\sc reproject}\footnote{\url{https://pypi.org/project/reproject/}}. The typical error in $I_{H\alpha}$ after adapting it to S-PASS is 0.6 R.

\subsection{Dust reddening}
The Galactic dust reddening towards Sh 2$-$27 was gathered from the Galactic Dust Reddening and Extinction map of \citet{Schlegel1998}, and corrected according to the updated estimates by \citet{Schlafly2011} (see Sect.~\ref{sec:em}). This map calculates the total dust reddening along the LoS. We used it to correct for the intervening dust extinction along the LoS at the coordinates of Sh 2$-$27. Similar to the SHASSA map, we smoothed and resampled the map as a final dust extinction map before combining it with other datasets.

 \section{Method}\label{sec:method}
As the 3D structure of the H\,{\sc ii} region has an impact on the understanding of its LoS-integrated polarization properties, we modelled the geometrical shape of Sh 2$-$27 as an ellipsoid to estimate the path length. In the following sections, we introduce our methods of using $I_{H\alpha}$ to estimate $n_{e}$ by calculating emission measure ($EM$), and using polarization angles to obtain $B_{\parallel}$ by estimating \textit{RM}. 
 
 \subsection{The ellipsoid model}\label{sec:ell_mod}
 To estimate the path length through the H\,{\sc ii} region (the LoS thickness), we modelled Sh 2$-$27 to be an ellipsoid, which matches its observed shape both in S-PASS and SHASSA.
 
 We applied the general ellipsoid equation rewritten as \(z=s\sqrt{1-\frac{x^2}{a^2}-\frac{y^2}{b^2}}\) in Cartesian coordinates and applied a basic coordinate transformation, where the semi-axes have lengths of $a$ (semi-major radius), $b$ (semi-minor radius), and $s$ (LoS radius or LoS thickness); $x$ and $y$ are in the plane of the sky; and $z$ is along the LoS. Even though the LoS thickness of the H\,{\sc ii} region is unknown, a reasonable estimate is that $s$ must be between the values of the plane-of-sky axes $a$ and $b$, hence $n_{e}$ and $B_{\parallel}$ are  zero for \(\frac{x^2}{a^2}+\frac{y^2}{b^2}>1\). We assume $a=s$ here, but we discuss the ramifications of this choice for $n_{e}$ and $B_{\parallel}$ in Sect.~\ref{sec:discussion}. 
 
\begin{figure}
  \resizebox{\hsize}{!}{\includegraphics{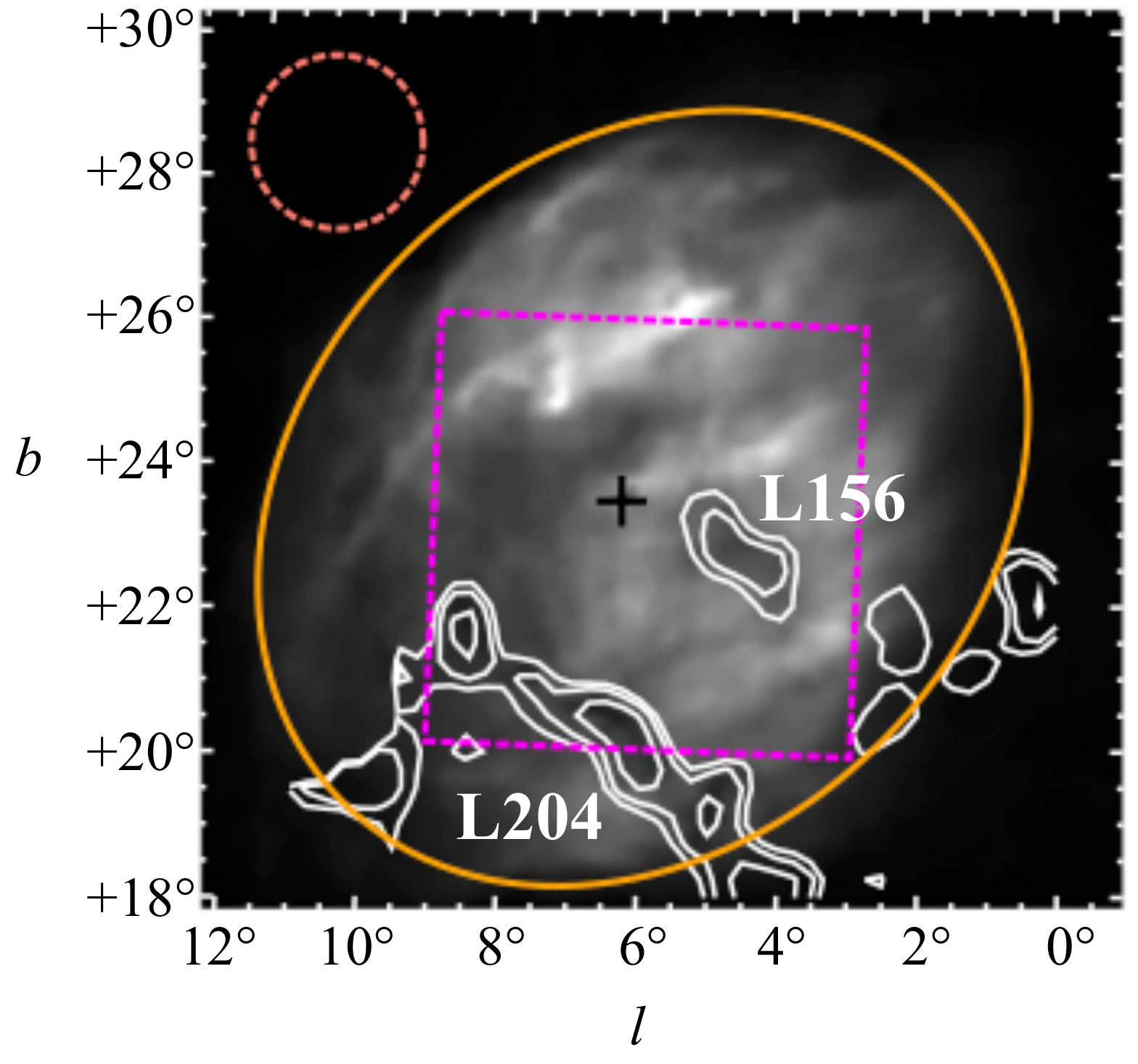}}
  \caption{Selected regions and the dark clouds (this work) overlaid on the $I_{H\alpha}$ map of Sh 2$-$27. The dashed circle with a radius of 1${\fdg}$2 is chosen to estimate the off-region used in $RM\textsubscript{Sh2--27}$ estimation (see Sect.~\ref{sec:rm}). The dashed box is used for the \textit{SF} calculations. Each side of the box is $\sim$6$^{\circ}$. Contours from the CO emission map of \citet{Dame2001} are overlaid on the H\,{\sc ii} region to display the structure of the dark clouds. The plus sign represents the location of $\zeta$~Oph at $(l, b)=(6{\fdg}3, +23{\fdg}6)$. The greyscale range is 0$-$2000 dR.}
  \label{fig:regions}
\end{figure}

\begin{figure}
  \resizebox{\hsize}{!}{\includegraphics{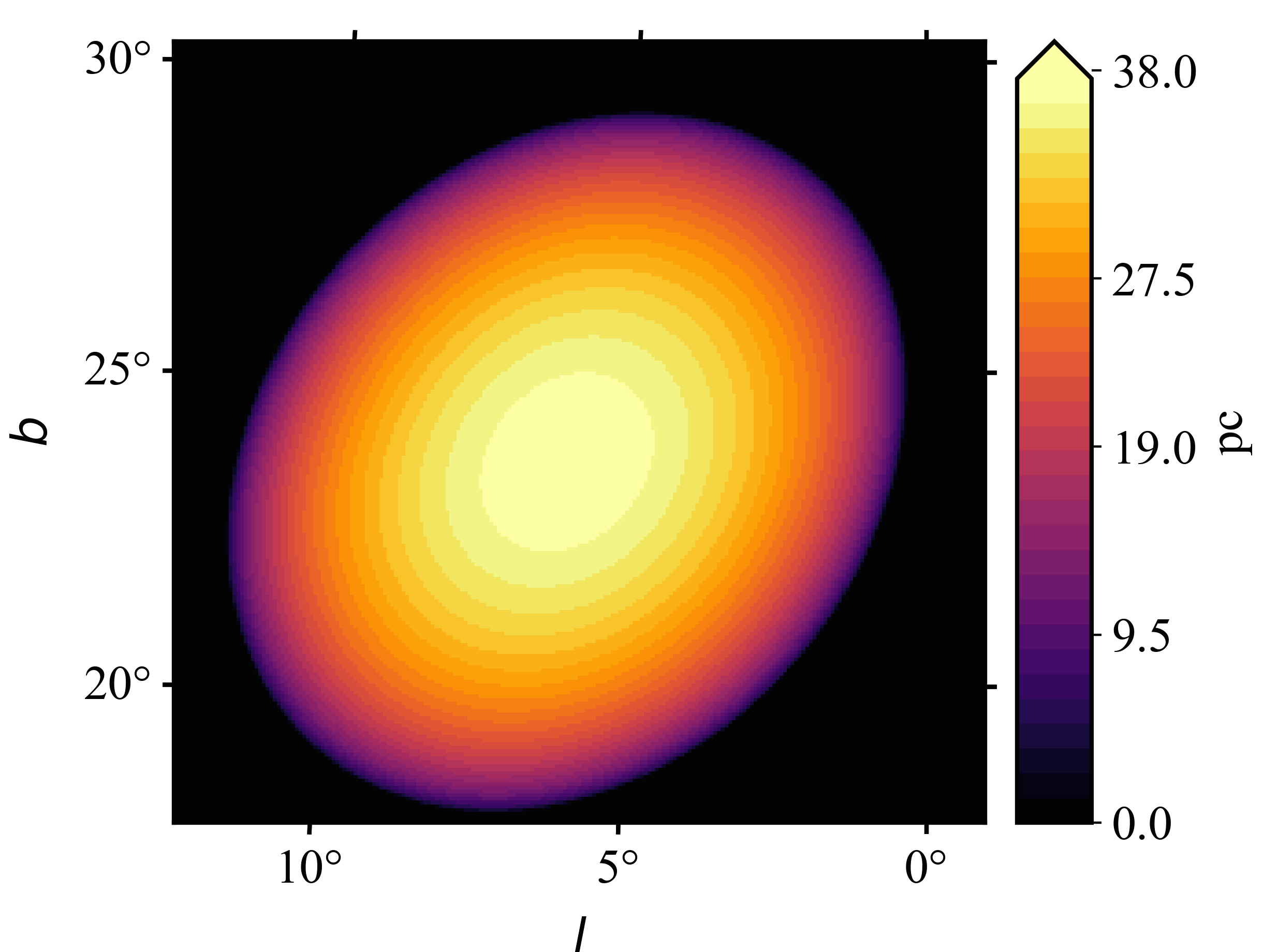}}
  \caption{Colour-coded LoS values for each pixel  represented by the projected elliptical shape chosen for Sh 2$-$27. The maximum path length through the H\,{\sc ii} region ($z$-axis) is equal to the major axis of the ellipsoid (38~pc).}
  \label{fig:los}
\end{figure}

 At the distance of 180 pc, the corresponding semi-minor axis is 15 pc and the semi-major axis is 19 pc. The resulting elliptical region can be seen in Fig.~\ref{fig:regions}. The ellipsoidal shape is geometrically obtained by revolving its surface about its major axis with a central position in $(l, b)=(6{\fdg}3, +23{\fdg}6)$, and a major axis inclination angle of $45{\fdg}8$ with respect to the direction of the Galactic longitude. In that way, $\zeta$~Oph is in the geometric centre of the H\,{\sc ii} region. The resulting LoS thickness values can be seen in Fig.~\ref{fig:los}. As expected from an ellipsoidal shape, the LoS values gradually decrease farther away from the centre.

\subsection{Emission measure}\label{sec:em}
The \textit{EM} is related to $n_{e}$ via the equation
\begin{equation}
\left(\frac{EM}{\rm cm^{-6}\,pc} \right) = \intop_{0}^{\infty} \left(\frac{n_e}{{\rm cm^{-3}}}\right)^{2} \left( \frac{ds}{\rm pc} \right).
\end{equation}
\textit{EM} can also be derived from the intensity of the H$\alpha$ line due to recombination transitions of neutral hydrogen atoms, according to \citep{Reynolds1988} :
\begin{equation}
\label{eq:EM}
\left(\frac{EM}{\rm cm^{-6}\,pc} \right) = 2.75~ \left(\frac{T_e}{10^{4}{\rm K}}\right)^{0.9}  \left(\frac{I_{H\alpha}}{{\rm R}}\right) e^{\tau}
.\end{equation}
In the following paragraphs we explain $T_{e}$, $I_{H\alpha}$, and $e^{\tau}$.

The electron temperature, $T_{e}$, is assumed to be $\sim$7000~K for this H\,{\sc ii} region\footnote{\citet{Nicholls2012} derived generally higher electron temperatures for H\,{\sc ii} regions if the electron velocities  followed a $\kappa$-distribution instead of a Maxwell-Boltzmann distribution. However, \citet{Draine2018} showed that a $\kappa$-distribution is not realistic in H\,{\sc ii} regions as this would relax to a Maxwell-Boltzmann distribution very quickly. Consequently, we use the estimation of $T_e = 7000$~K. \citep{Reynolds1982,Reynolds1988}. \citet{Madsen2006} also showed that the H\,{\sc ii} regions ionized by characteristic O-type stars have temperatures between 6000~K and 7000~K.}.

The surface brightness values, $I_{H\alpha}$, from SHASSA are in deci-Rayleigh (dR), and Fig.~\ref{fig:emission}(a) shows the H$\alpha$ map of Sh 2$-$27. Next to the patches of bright H$\alpha$ emission, the most conspicuous structures are filaments of low H$\alpha$ emission towards the south of the image. These low-intensity features are caused by the absorption of H$\alpha$ emission by the dark clouds in front of or partially overlapping with Sh 2$-$27 \citep{Tachihara2000a,Choi2015}. Figure~\ref{fig:regions} shows the projected locations of these clouds in CO emission \citep{Dame2001}. As seen in Fig.~\ref{fig:emission}(b), the high dust reddening in these clouds is rather pronounced. As the anti-correlation between the H$\alpha$ emission and the dust extinction indicates, some H$\alpha$ emission throughout Sh 2$-$27 is also absorbed by the intervening dust. 

We correct for the dust extinction, $e^{\tau}$, by including the optical depth ($\tau$) in the computation of \textit{EM} in Eq.~\ref{eq:EM} as follows. To correct for the intervening dust along the LoS, we include an optical depth estimation, \(\tau=2.44 \times E(B-V)\) \citep{Finkbeiner2003}. The  reddening in magnitudes $E(B-V)$ is given in the dust reddening map of \citet{Schlegel1998}, corrected by the  \citet{Schlafly2011} term as \(E(B-V)_{S\&F} = 0.86 \times E(B-V)_{SFD}\), where $S\&F$ symbolizes the  \citet{Schlafly2011} values derived by \citet{Schlegel1998} values (\textit{SFD}). Here we assume that the dust is in front of the H\,{\sc ii} region and is not mixed in. The maps of $E(B-V)$ and $\tau$ are presented in Figs.~\ref{fig:emission}(b) and \ref{fig:emission}(c), respectively, and as seen, the dust reddening is higher towards the dark clouds, especially towards L204. 

Based on the simplifying assumptions on the extinction correction, the absorption by the foreground dark clouds,   clearly visible in Fig.~\ref{fig:emission}(a), has lessened after dust correction. Likewise, the imprint of the dust map on the \textit{EM} map has diminished significantly (see Fig.~\ref{fig:emission}d). We assume that the effect of the dust extinction on the \textit{EM} map is negligible after correction. Moreover, we do not include any background or foreground H$\alpha$ emission correction as we assume that  the \textit{EM} signal is caused by the H\,{\sc ii} region itself due to the lack of the background \textit{EM} signal surrounding Sh 2$-$27, as seen in Fig.~\ref{fig:emission}(d).

\subsection{Electron density}\label{sec:ne}
\begin{figure*}
\centering
   \includegraphics[width=17cm]{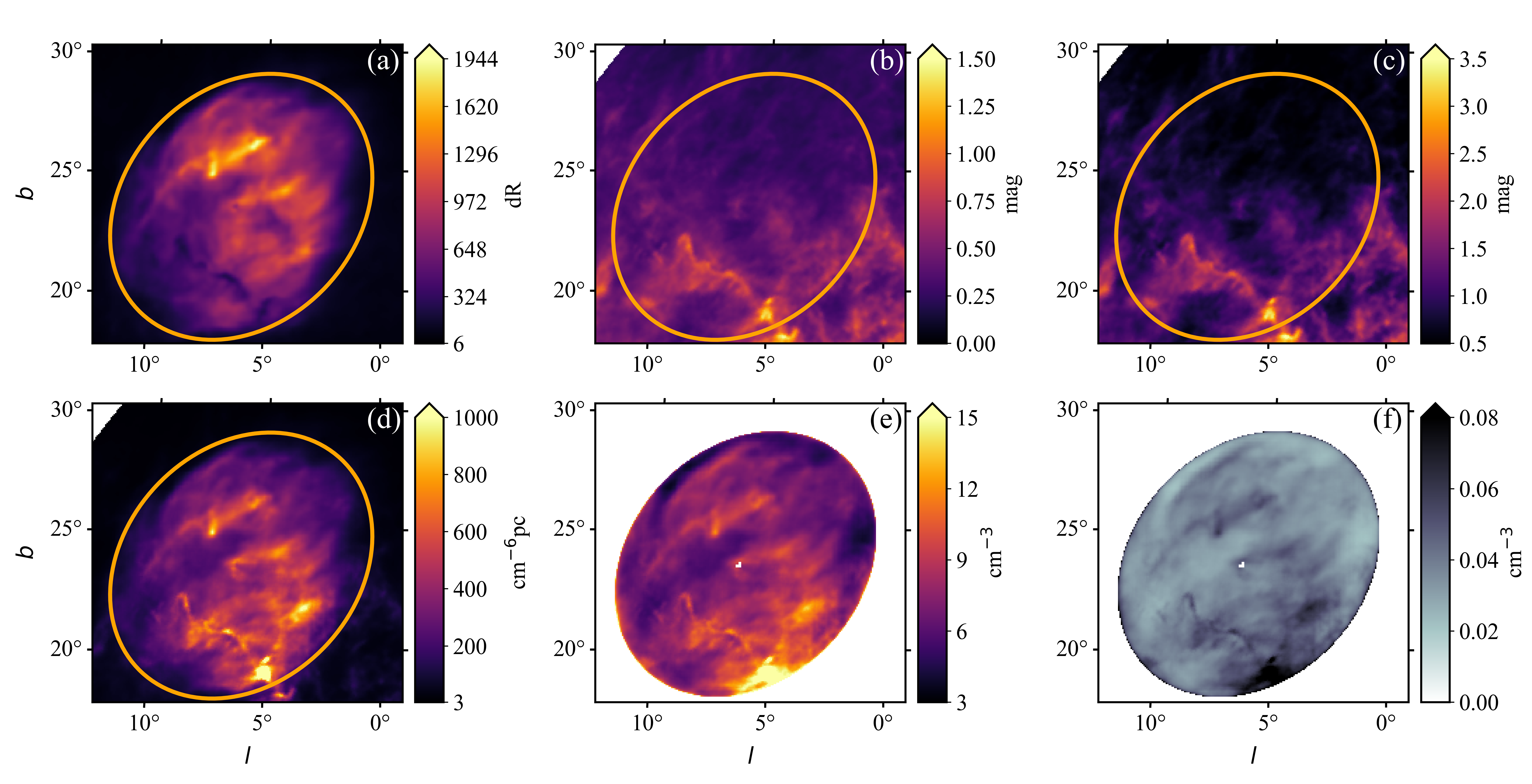}
  \caption{Efficacy of the extinction correction, as described in Sect.~\ref{sec:method}. (a) $I_{H\alpha}$; (b) $E(B-V)$; (c) $\tau$; (d) $EM$ (colour-scale  saturated to 1000 cm$^{-6}$~pc, the maximum is 2251 cm$^{-6}$~pc); (e) $n_{e}$ (colour-scale  saturated to 15 cm$^{-3}$, the maximum is 57.6 cm$^{-3}$); (f) $\sigma_{n_{e}}$ (colour-scale  saturated to 0.09 cm$^{-3}$, the maximum is 0.29 cm$^{-3}$). The orange elliptical region gives the area chosen for the LoS estimations. In (e) and (f), $\zeta$~Oph is masked to prevent confusion in the colour-scale. Throughout the paper, the uncertainties are shown in greyscale.}
  \label{fig:emission}
\end{figure*}

\begin{figure}
  \resizebox{\hsize}{!}{\includegraphics[width=8.5cm]{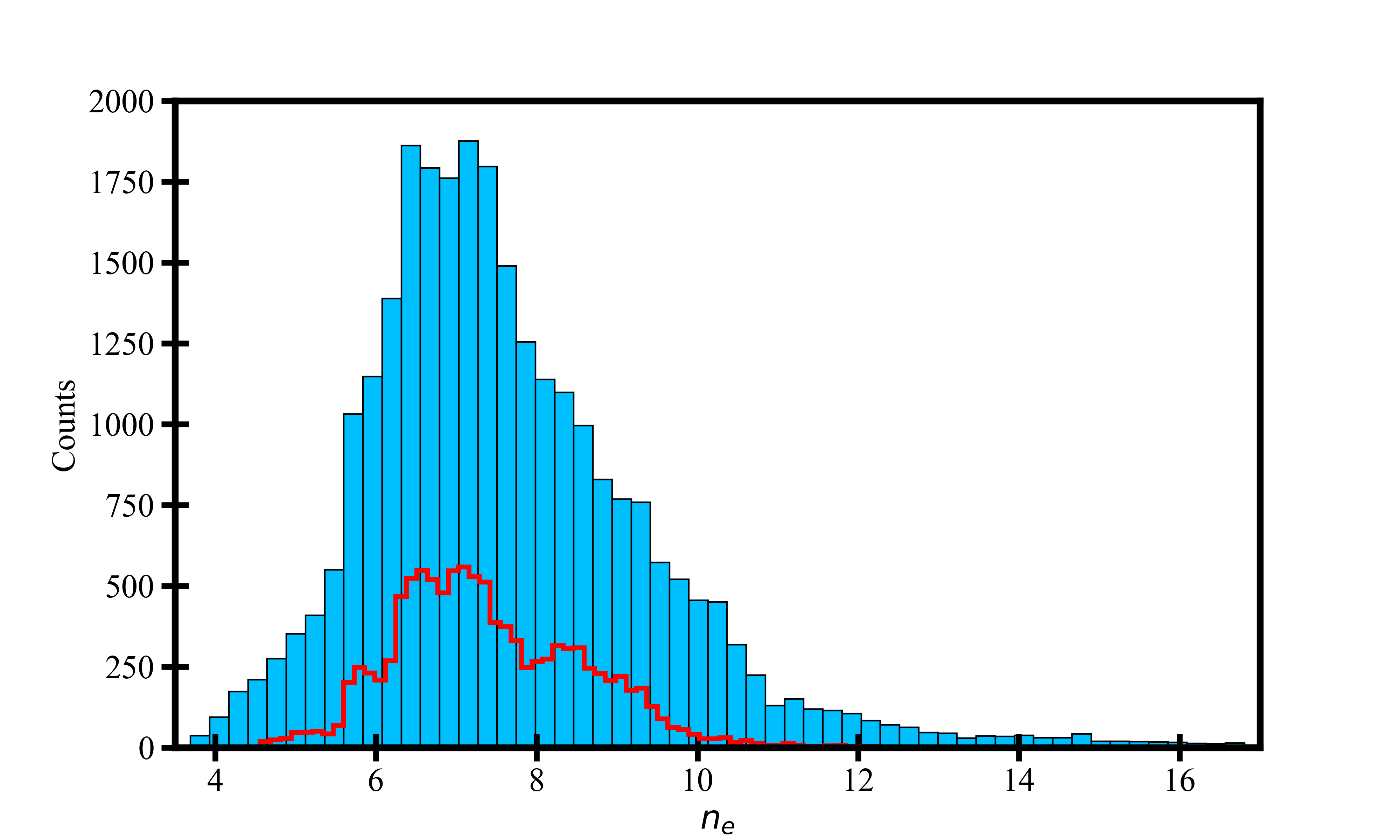}}
  \caption{Distribution of  $n_{e}$  in the elliptical area  selected for Sh 2$-$27. The overlaid red histogram shows the distribution gathered from the box region shown in Fig.~\ref{fig:regions}.}
  \label{fig:ne_hist}
\end{figure}

Since the ionized medium of an H\,{\sc ii} region is inhomogeneous, the term $f$ stands for a fraction of the volume of dense gas (clumps) to a total volume of an H\,{\sc ii} region that is occupied by the clumps. Different approaches to estimate $f$ in  H\,{\sc ii} regions can be found in earlier studies (e.g. \citealt{Herter1982,Kassim1989,Giammanco2004}). To account for inhomogeneities of the ISM through the H\,{\sc ii} region, we assumed $f$ of the Faraday-rotating gas to be 0.2 \citep{Harvey-Smith2011}, and outside a clump $n_{e}$ is zero. Equation~\ref{eq:ne} can be  used to evaluate the variations in the electron density averaged along the LoS as \citep{Ocker2020} 
\begin{equation}
\label{eq:ne} 
\left(\frac{n_{\rm e}}{\rm cm^{-3}}\right) = \left( \frac{EM}{\rm cm^{-6}\,pc} \right)^{1/2} \left( \frac{f s}{\rm pc} \right)^{-1/2}.
\end{equation}
We found a median value of $n_{e}$ ($\overline{n}_{e}$) inside the elliptical area to be $7.3\pm0.1$~cm$^{-3}$, and the resulting map can be seen in Fig.~\ref{fig:emission}(e); for the map of $\sigma _{n_{e}}$ see Fig.~\ref{fig:emission}(f). The high $n_{e}$ values in the extension of L204 at $(l, b) \sim (5^{\circ}, +19^{\circ})$ show the highest dust (Fig.~\ref{fig:emission}b) and the highest uncertainties (Fig.~\ref{fig:emission}f). These values are likely caused by a possible imperfect correction for the dust extinction. Nevertheless, this region falls outside the area where we perform the quantitative analysis of the turbulence (Sect.~\ref{sec:turbulence}); thus, we do not attempt any further corrections. Figure~\ref{fig:ne_hist} shows the distribution of $n_{e}$, as well as the distribution of $n_{e}$ found inside the box in Fig.~\ref{fig:regions}.

\section{Analysis of the polarized radio observations}\label{sec:analysis_interp}
In what follows we briefly introduce the total radio and the polarized radio intensities and describe how we estimate the polarization angles that we used to calculate $RM\textsubscript{Sh2--27}$ to derive a map of $B_{\parallel}$.

\subsection{Total radio and polarized radio intensities in Sh 2$-$27}\label{sec:pol}
\begin{figure}
  \resizebox{\hsize}{!}{\includegraphics[width=6.7cm, height=8.9cm]{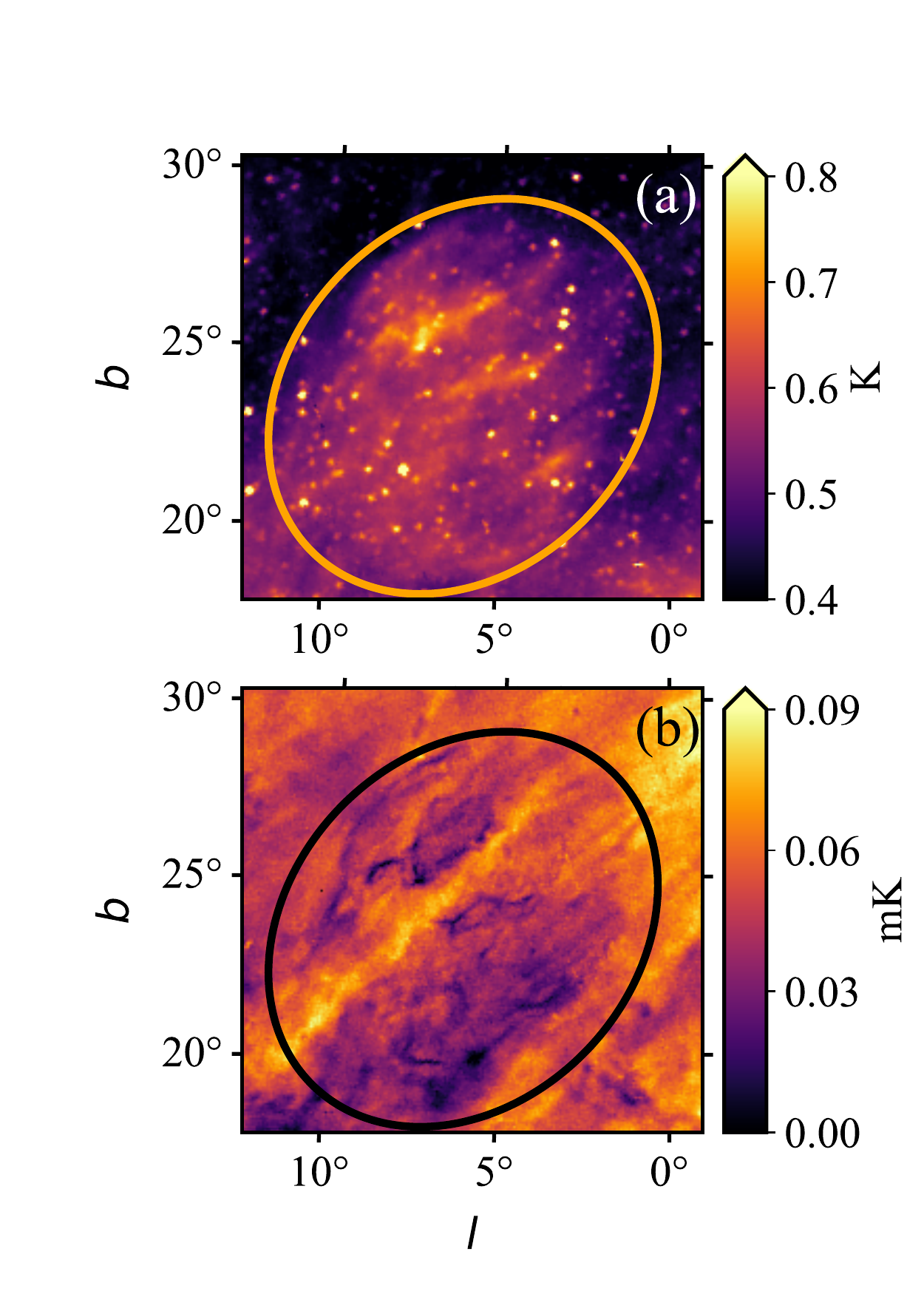}}
  \caption{Radio synchrotron emission in the direction of Sh 2$-$27. Top: Total radio intensity (colour-scale saturated to 0.80 K, the maximum is 0.95 K). Bottom: Polarized radio intensity.}
  \label{fig:synch}
\end{figure}

Figure~\ref{fig:synch} presents the total radio and polarized radio intensities in Sh 2$-$27. The total intensity denotes a combination of free-free emission and synchrotron emission, while the polarization only occurs in synchrotron emission. In total intensity the free-free emission may come from where the Stokes $I$ map correlates with the H$\alpha$ map. In addition, there are three noticeable features in the polarized intensity, which we will discuss in turn: a) the polarized intensity does not show emission correlating with the total intensity;
b) depolarization canals are visible; c) there is a highly polarized filament diagonally crossing the H\,{\sc ii} region.

Firstly, if there were significant synchrotron radiation emitted by Sh 2$-$27, there would be accompanying polarized emission. As this is not seen in Fig.~\ref{fig:synch}, synchrotron emission from Sh 2$-$27 is negligible, supporting the assumption that Sh 2$-$27 is a Faraday screen, as discussed in Sect.~\ref{sec:sh2-27}. 

Secondly, depolarization canals are elongated structures of one beam-width  depolarization, without counterparts in total radio intensity (e.g. \citealt{Haverkorn2000,Gaensler2001}). They are ubiquitous in radio polarization maps (e.g. \citealt{Gray1999, Uyaniker1999, Haverkorn2000, Gaensler2001, Shukurov2003, HaverkornHeitsch2004, Fletcher2007}). Depolarization canals can be caused by beam depolarization and depth depolarization \citep{Fletcher2007}. Beam depolarization occurs if structures in the Faraday rotating medium give rise to significant gradients in polarization angle within the telescope beam. Depth depolarization occurs when a medium along a LoS is mixed with both a synchrotron-emitting medium and a thermal medium, which causes different polarization angles at different depths in the medium, reducing the observed polarization. We note that this is not the case for Sh 2$-$27, which does not produce a significant amount of polarized emission itself (see Sect.~\ref{sec:sh2-27}). The observed depolarization canals point towards sharp gradients in Faraday rotation in the H\,{\sc ii} region within the S-PASS beam, as might be expected from small-scale structures in the magnetic field and/or electron density.

\begin{figure*}
\center
  \includegraphics[width=12cm]{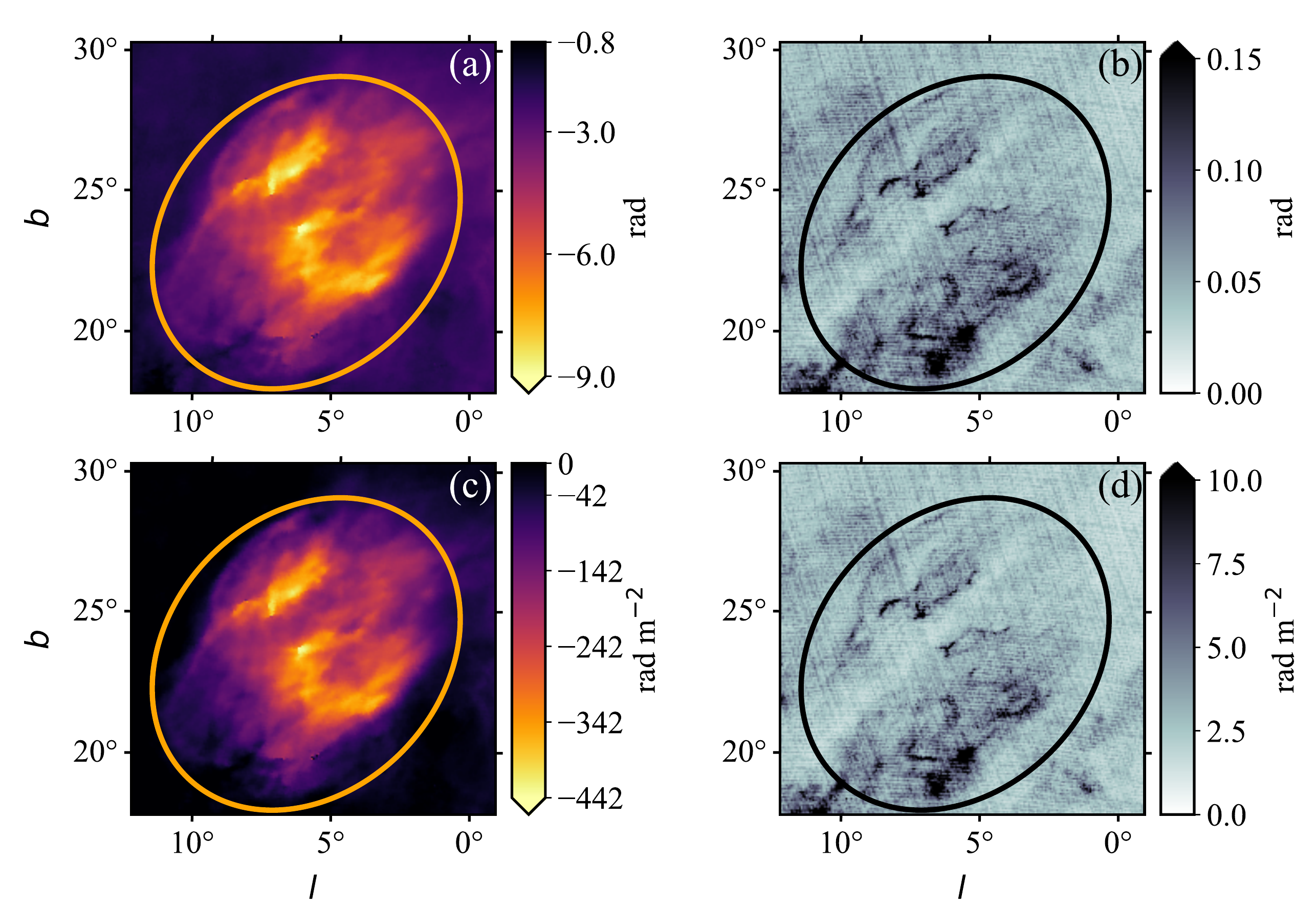}
  \caption{Polarization angle and the rotation measure properties of Sh 2$-$27. (a) $\chi$; (b) $\sigma_{\chi}$ (colour-scale  saturated to 0.15 rad, the maximum is 0.70 rad); (c) $RM\textsubscript{Sh2--27}$; (d) $\sigma_{RM\textsubscript{Sh2--27}}$ (colour-scale  saturated to 10 rad~m$^{-2}$, the maximum is 45 rad~m$^{-2}$). The orange and black ellipses show the area dealt with in the calculations.}
  \label{fig:radio}
\end{figure*}

Lastly, the polarized filament crossing Sh 2$-$27 is not part of the H\,{\sc ii} region, but belongs to a larger configuration of radio-filaments and loops. It can be seen in the WMAP K-band polarization maps of \citet{Vidal2015}, where it appears between the filament IX and the Galactic Centre spur (in their Fig.~2), and extends towards the north Galactic pole along with the large-scale magnetic field direction (in their Fig.~1 panel top left). It is not visible either in the lower-frequency radio polarization maps of \citet{Thomson2019} in 300$-$480 MHz or in the 1.4 GHz all-sky polarization map combined by \citet{Reich2009}. It might be unresolvable because of the low resolution. As there are no depolarization canals across the polarized filament, it has to be in the foreground of the H\,{\sc ii} region. The non-zero polarized intensity of the filament contributes to the background intensity, and therefore can still influence the observed polarization angles \citep{Kumazaki2014}, as well as the \textit{RM} measurements. There may be a hint of the filament in the \textit{RM} maps (see Fig.~\ref{fig:radio}). However, as these data do not allow quantification or correction of this slight influence, we  assume that it is negligible.

\subsection{Polarization angle}\label{sec:pa}
H\,{\sc ii} regions as Faraday screens alter the polarization angle of a traversing electromagnetic wave on the plane of the sky. The polarization angle, $\chi$, can be calculated by using Stokes $Q$ and $U$ :
\begin{equation}
 \left(\frac{\rm \chi}{\rm rad} \right) = \frac{1}{2}\, \rm{arctan}\left(\frac{\rm U}{\rm Q} \right). 
 \label{eq:PA}
\end{equation}
Following the estimation of the polarization angle, we de-rotated the relevant data in n$\pi$-ambiguity between adjacent data points ($i,j$) to check if a jump in the angle occurred,  caused by a polarization vector that revolved multiple times ($n$). Hence, we determined the most probable angle relative to a surrounding reference point in the consideration of \(\Delta \chi=\chi_{(i+1,j)} - \chi_{(i,j)}\) \citep{Brown2003,Haverkorn2003a,Brentjens2005}. The de-rotation is made according to the following equation:
\begin{equation}
\chi _{(i,j)} = \begin{cases}
\chi _{(i+1,j)}, & -\frac{\pi}{2} \leq \Delta \chi \leq \frac{\pi}{2} \\
\chi _{(i+1,j)}-\pi, & \Delta \chi > \frac{\pi}{2} \\
\chi _{(i+1,j)}+\pi, &\Delta \chi < -\frac{\pi}{2} \\
\end{cases}
\label{npi}
\end{equation}
We note that employing this algorithm row by row may result in the $n\pi$-ambiguity being unresolvable in some sequential pixels (only 6 out of 47073 pixels), which are centred at the following locations (they are left uncorrected): $(l, b)=(5{\fdg}5, +19{\fdg}9)$, $(5{\fdg}5, +20{\fdg}0)$, $(5{\fdg}7, +19{\fdg}8)$, $(5{\fdg}7, +19{\fdg}9)$, $(8{\fdg}3, +25{\fdg}3)$, $(8{\fdg}4, +25{\fdg}4)$. These pixels are in the highest depolarization regions seen in Fig.~\ref{fig:synch}(b). The resulting map can be seen in Fig.~\ref{fig:radio}(a) along with its uncertainty for each data point in Fig.~\ref{fig:radio}(b). Additionally, as seen in Fig.~\ref{fig:radio}(b), $\sigma _{\chi}$ resembles the reverse of Fig.~\ref{fig:synch}(b), since \(\sigma _{\chi} \propto P^{-1}\) in  Eq.~A.12 of \citealt{Brentjens2005}.

\begin{figure}
  \resizebox{\hsize}{!}{\includegraphics{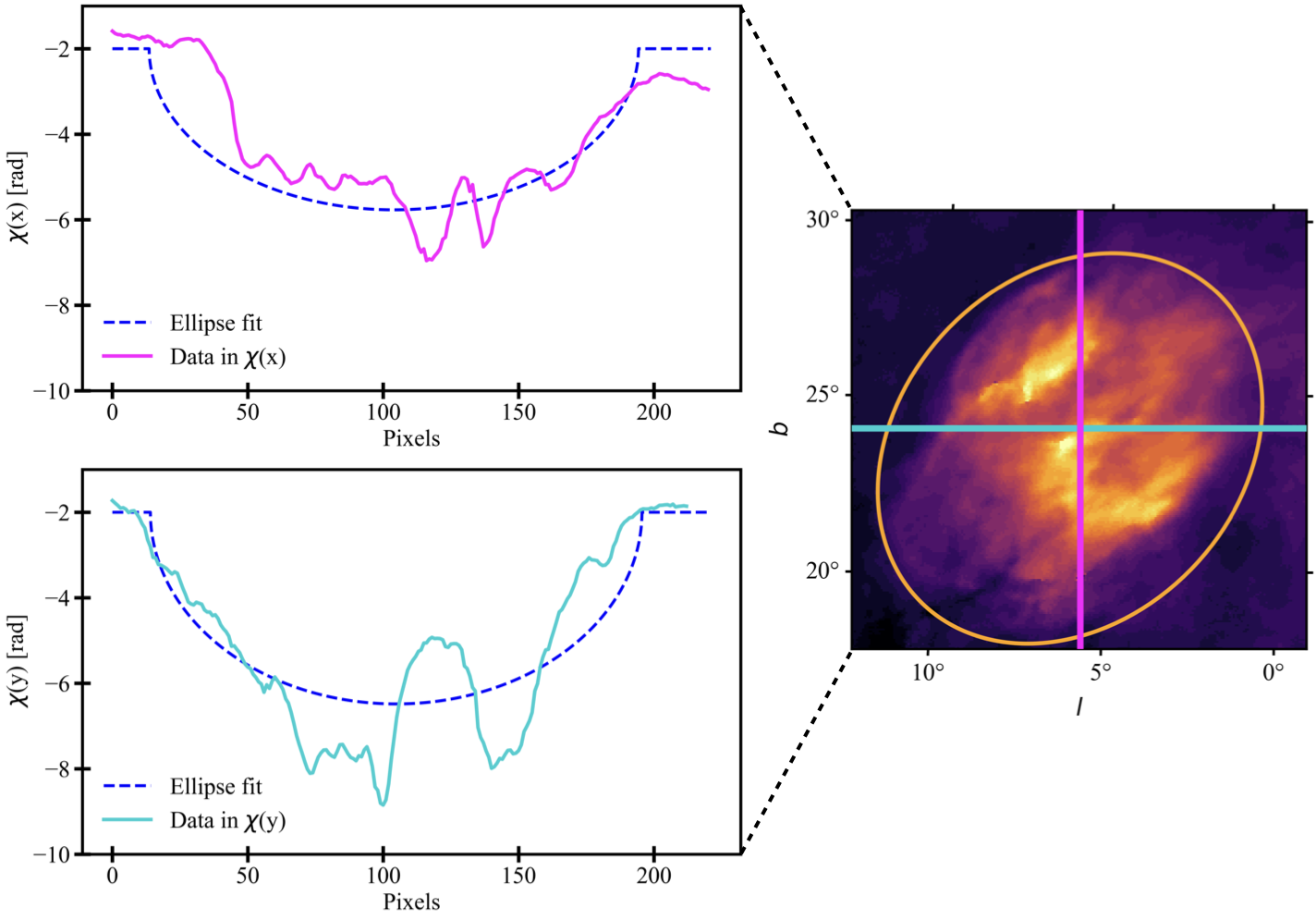}}
  \caption{Comparison of the elliptical model to the polarization angle variations in Sh 2$-$27. Left: n$\pi$-ambiguity corrected polarization angle variations along 1D image slices. Right: Colour-coded lines show the data taken from the polarization angle map. A slice of the data in the $xy$-plane is considered, passing through $l=6\degr$ for $\chi(x)$ and $b=23\degr$ for $\chi(y)$, where $(l,b)=(6\degr, 23\degr)$ is the centre of the ellipse.}
  \label{fig:ellipse_fit}
\end{figure}

To see how the elliptical area that we choose is related to the polarization angle variations, and to  the  $I_{H\alpha}$ variations, we show a basic model fitting in Fig.~\ref{fig:ellipse_fit}. This figure shows cuts from the H\,{\sc ii} region on both axes of the polarization angle map (Fig.~\ref{fig:radio}a). The ellipsoid parameters that we mention in Sect.~\ref{sec:ell_mod} are fitted to these data. They roughly demonstrate how reasonably the size of the ellipse is determined for the maximum LoS thickness of the H\,{\sc ii} region, which is essential for $n_{e}$ and $B_{\parallel}$ estimations. Consequently, they can be interpreted as small-scale fluctuations in the magnetic field and/or the electron density, as also indicated by the presence of the depolarization canals.

\subsection{Rotation measure}\label{sec:rm}
For a single-frequency polarization measurement of Sh 2$-$27, we can estimate $RM\textsubscript{Sh2--27}$ as long as we can treat the H\,{\sc ii}
 region as a Faraday screen (as shown in e.g. \citealt{Iacobelli2014,Robitaille2017,Robitaille2018}, and in Sect.~\ref{sec:pol}). Here, we initially assume that the background and foreground \textit{RM}s are constant \citep{Sun2007,Xiao2011,Thomson2018}. This is a reasonable assumption relative to the high \textit{RM} in the H\,{\sc ii} region itself. Each pixel within the elliptic boundary of Sh 2$-$27 holds that the polarization angle
\begin{equation}
\begin{split}
  \chi_{on}(x,y) & =  RM(x,y) \lambda_{S}^2 +\chi_0 \\
   & = \left(RM\textsubscript{Sh2--27}(x,y) + RM_{back}\right) \lambda_{S}^2 +\chi_0,
   \end{split}
\end{equation}
where $\lambda_S = 0.13$~m is the observing wavelength, $RM\textsubscript{Sh2--27}(x,y)$ is the position-dependent $RM\textsubscript{Sh2--27}$ of the H\,{\sc ii} region itself, $RM_{back}$ is the position-independent background \textit{RM}, and $\chi_0$ is the intrinsic polarization angle assumed to be constant. The polarization angle just off the H\,{\sc ii} region is  then  
\begin{equation}
  \chi_{off} = RM_{back} \lambda_{S}^2 + \chi_0;
\end{equation}
hence, RM$\textsubscript{Sh2--27}$ can be calculated as
\begin{equation}
  RM\textsubscript{Sh2--27}(x,y) = \frac{\chi_{on}(x,y) - \chi_{off}}{\lambda_{S}^2},
\end{equation}
where \(\chi_{off} = -1.8\) rad. By choosing $\chi_{off}$ as the polarization angle value just outside the ellipsoid, which is de-rotated to the value closest to the polarization angle at the edge (as in Sect.~\ref{sec:pa}), we avoid the n$\pi$-ambiguity problem here, as $\chi_{on}$ and $\chi_{off}$ individually have an n$\pi$-ambiguity, but \(\chi_{on}-\chi_{off}\) does not.

The off-region was chosen as the region close to the edge of Sh 2$-$27 with the least contaminating emission from other sources. Firstly, on the right side of Fig.~\ref{fig:radio}(a) there is  another H\,{\sc ii} region (Sh 2$-$7)  (see \citealt{Iacobelli2014}). Secondly, an off-region from the bottom left part of Fig.~\ref{fig:radio}(a) is also not ideal, considering the high errors in the polarization angle (Fig.~\ref{fig:radio}b). However, the off-region as drawn in Fig.~\ref{fig:regions} seems to be representative because, if we vary the radius or position of the circle in the vicinity, the off-region level does not vary significantly.

The resulting $RM\textsubscript{Sh2--27}$ map together with its uncertainty can be seen in Figs.~\ref{fig:radio}(c) and \ref{fig:radio}(d), respectively. The median $RM\textsubscript{Sh2--27}$ inside the elliptical area is $-126.0\pm3.1$~rad~m$^{-2}$. We estimated $\sigma_{RM\textsubscript{Sh2--27}}$ using Eq.~\ref{eq:ap_sigmarm}. This approach differs slightly from that of \citet{Harvey-Smith2011}, who used \textit{RM}s of polarized extragalactic point sources from the catalogue of \citet{Taylor2009}, and subtracted a small linear gradient in extrinsic \textit{RM} computed from a region around Sh 2$-$27. Their |\textit{RM}| values are roughly in the  range from 75 to 300~rad~m$^{-2}$, and ours in the range of 59~rad~m$^{-2}$ < |$RM\textsubscript{Sh2--27}$| < 443~rad~m$^{-2}$ within the elliptical boundary of the H\,{\sc ii} region. These values are generally consistent with each other, although our |$RM\textsubscript{Sh2--27}$| values show a higher maximum than their results. These high values are in regions where \citet{Harvey-Smith2011} do not have coverage of background sources, which makes it impossible to detect these values with their method.

\subsection{Magnetic field parallel to the LoS}\label{sec:B}
\begin{figure}
  \resizebox{\hsize}{!}{\includegraphics{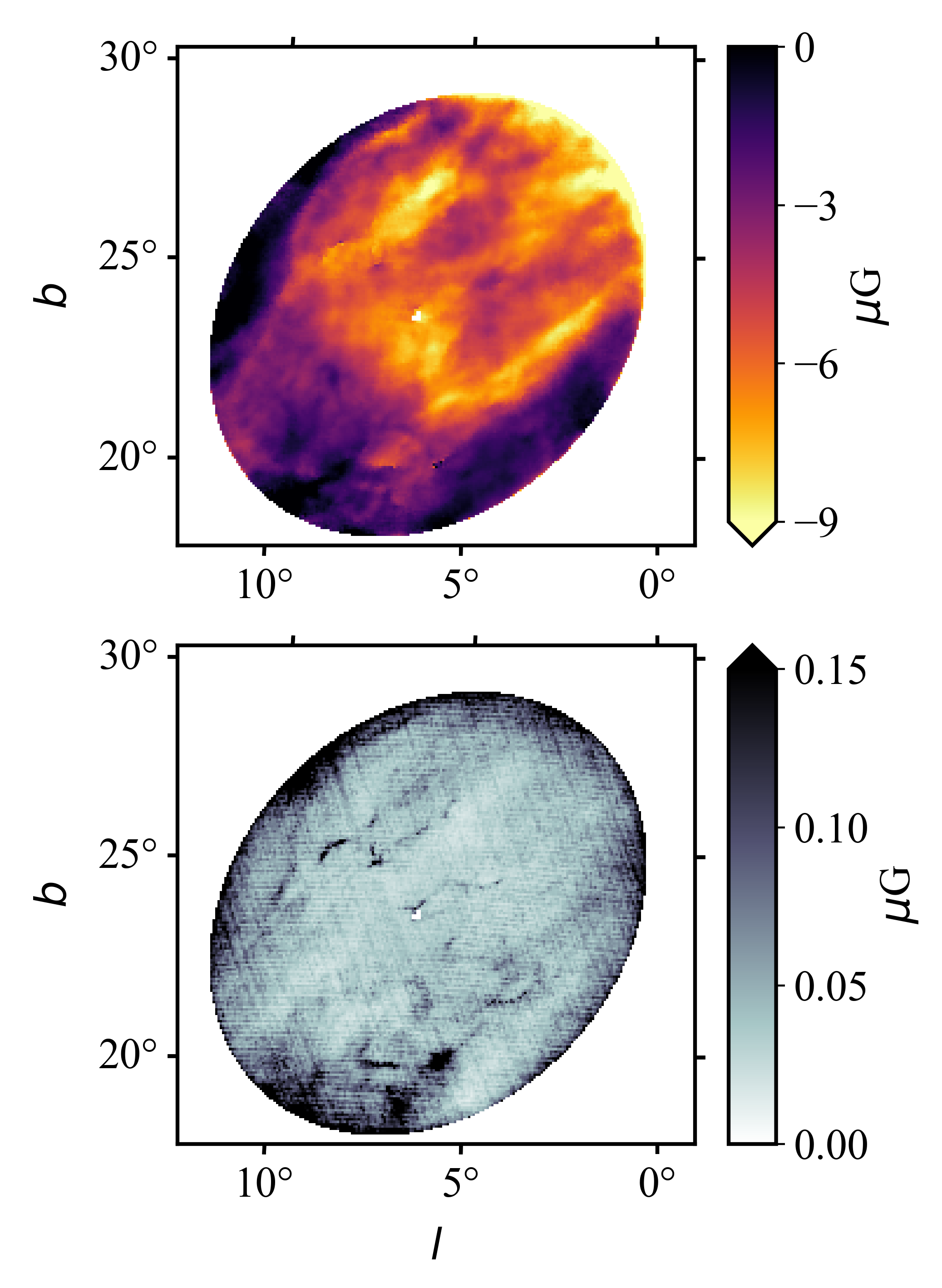}}
  \caption{$B_{\parallel}$ estimated in Sh 2$-$27 (top), and its uncertainty (bottom) in the ellipsoid path-length model of Sh 2$-$27. The colour-scale of $B_{\parallel}$ is limited between $-$9 $\mu$G and 0 $\mu$G; the minimum is around $-$18 $\mu$G and the maximum is around 3.5 $\mu$G. The colour-scale of $\sigma_{B_{\parallel}}$ is saturated to 0.15 $\mu$G; the maximum is $\sim$4.2 $\mu$G.}
  \label{fig:mf}
\end{figure}

We derived $B_{\parallel}$ for each pixel (i.e. for each LoS) using Eq.~\ref{eq:B}, which is rewritten from Eq.~\ref{eq:RM} under the assumption that $n_{e}$ and $B_{\parallel}$ are constant along the LoS:
\begin{equation}
\label{eq:B} 
 \left(\frac{B_{\rm ||}}{\rm \mu G}\right) = \frac{\rm 1}{\rm 0.812} \left( \frac{ RM\textsubscript{Sh2--27}}{\rm rad\,m^{-2}} \right) \left( \frac{ EM}{\rm cm^{-6}\,pc} \right)^{-1/2} \left( \frac{f s}{\rm pc} \right)^{-1/2}
.\end{equation}
This equation is also strictly valid if $n_{e}$ and $B_{\parallel}$ are uncorrelated. Figure~\ref{fig:mf} displays the result of $B_{\parallel}$ along with its uncertainty ($\sigma_{B_{\parallel}}$) using the relation in Eq.~\ref{eq:ap_sigmaB}.
We estimated $B_{\parallel}$ with a median ($\overline{B}_{\parallel}$) of $-4.5\pm0.1$~$\mu$G inside the elliptical area. We note that the extreme $B_{\parallel}$ values at the edges of the ellipse are likely caused by overestimation due to the small LoS values.

\section{Statistical analysis of turbulence with structure functions}\label{sec:turbulence}
The energy of incompressible and homogeneous turbulence cascades from the largest scales, where energy is deposited to the smallest scales through a dissipation process. If there is no energy input or loss in intermediate scales, this is called Kolmogorov turbulence \citep{Kolmogorov1941a}, which represents the turbulence with a basic scaling relation. This scaling is a power law, which can be characterized by a power spectrum (\textit{PS}) or a structure function (\textit{SF}). The theory of \citet{Kolmogorov1941a} predicts a power spectrum of 3D turbulence with a power-law slope of \(-11/3\). Likewise, \textit{SF} associated with the predicted 3D Kolmogorov turbulence is scaled by the power-law slope of \(5/3\) (for a comprehensive summary, see e.g. \citealt{Thompson2001}, Table 13.2). In 2D, power-law indices of \textit{PS} and \textit{SF} are also expected to be consistent with 2D Kolmogorov turbulence with a scaling of \(-8/3\) and \(2/3\) in the magnetized ISM, respectively \citep{Goldreich1995,Minter1996,Haverkorn2004-1,Haverkorn2006b,Mao2010}. We note that for other theories of turbulence, the relation between the power-law slopes of \textit{PS} and \textit{SF} may be different (e.g. \citealt{Cho2009}). Structure functions rise as fluctuations increase with larger scales. The scale where a structure function starts to flatten out is the largest scale. 

Since \textit{PS} and \textit{SF} give the same information \citep{Thompson2001}, to ease comparison with earlier studies we present only \textit{SF}s here. We also calculated \textit{PS} of our data, and it gave equivalent results in the slopes. The reliability of \textit{SF} requires a high-resolution observation as discussed in the study of \citet{Lee2016}, which is the case for S-PASS.

\subsection{Structure function}\label{sec:sf}
We estimate \textit{SF}s as follows. The angular separation between two source components is binned with equal lags in logarithmic intervals of a separation by averaging the squares of the differences between the pairs, that is, a radial averaging was made to calculate a 1D \textit{SF} from the 2D \textit{SF}. Thereby, a second-order \textit{SF} is calculated as
\begin{equation}
\label{eq:SF}
\mathit{SF_{g}}(\delta \boldsymbol{r_{n}})=\left\langle\left|\mathit{g}\left(\boldsymbol{r}\right)-\mathit{g}\left(\boldsymbol{r}+\delta \boldsymbol{r_{n}}\right)\right|^{2}\right\rangle_{\mathbf{\boldsymbol{r}}},
\end{equation}
where \(\textbf{\textit{r}}=(x,y)\) denotes a 2D position vector on the plane of the sky;  $\delta\textbf{\textit{r}}_{n}$ is a certain angular separation between the two points; \(\langle...\rangle_{\mathbf{\boldsymbol{r}}}\) implies an ensemble averaging over all positions with $\delta\textbf{\textit{r}}_{n}$ in the relevant interval; and  \textit{g} stands for any function of a field, in our case $B_{\parallel}$. We binned these data into bins of 25 pixels. Assuming Gaussian noise, we subtracted the error of the structure function of the noise (\textit{SF}$_{\sigma}$) from \textit{SF}$_{B_{\parallel}}$, as done in \citet{Haverkorn2004-1} and \citet{Stil2011}.

One issue to take into consideration in the calculation of \textit{SF}$_{B_{\parallel}}$ is that the pixel sizes in longitude and latitude are not equal due to the projection, which might be offset by including a factor cos(latitude) to the longitude coordinate. As Sh 2$-$27 is centred at a latitude of $\sim$23$\degr$, the distortion is not as extreme, but it is expanded by $\sim$9\% with respect to the latitude. A correction for this effect is included in our estimations, although it did not make any significant change in the \textit{SF}$_{B_{\parallel}}$ slopes.

The \textit{SF}$_{B_{\parallel}}$ is determined within the box region in Fig.~\ref{fig:regions}. The resulting \textit{SF}$_{B_{\parallel}}$ can be seen in Fig.~\ref{fig:SFs}. We note that the error bars of \textit{SF}$_{B_{\parallel}}$ are smaller than the plot symbols. Because small-scale structures are smoothed out below the beam-size, \textit{SF}$_{B_{\parallel}}$ is reliable down to the S-PASS resolution (8$\farcm$9). Additionally, in a box inside the H\,{\sc ii} region's projected area, \textit{SF}$_{B_{\parallel}}$ is reliable up to about half the box, that is, below $\sim$180$\arcmin$. Consequently, any fluctuations on   scales larger than this are unreliable.

The \textit{SF}$_{B_{\parallel}}$ shows a power-law behaviour with a power-law slope of 1.4 that fits the data between 10$\arcmin$ and 50$\arcmin$. On scales larger than that the power-law slope decreases to a flat \textit{SF}$_{B_{\parallel}}$ on a scale of $\sim$10~pc, which corresponds roughly to half of the box. On the largest scales, \textit{SF}$_{B_{\parallel}}$ seems to turn up again, which is an artefact due to the poor sampling of very large scales in the box. It is tempting to interpret the turnover scale of $\sim$10~pc as the outer scale of the turbulence. However, since this scale is comparable to the size of the box, we cannot be sure whether the turnover is due to the finite box size. The power-law slope of 1.4 is indeed higher than the Kolmogorov slope. Hence, to test the dependency of the slope of the estimated turbulent $B_{\parallel}$ on the actual magnetic field \textit{SF}$_{B_{\parallel}}$, and to investigate whether the turnover scale is due to the outer scale of the turbulence or due to the box size, we designed simulations of an ellipsoidal H\,{\sc ii} region containing an input $n_{e}$ and an input turbulent magnetic field. We describe the details of these simulations in the next section.

\subsection{Synthetic models}\label{sec:sim}
Synthetic models with Kolmogorov scaling can give us an idea of the observed turbulence in the presence of Sh 2$-$27. In that sense, introducing different outer scales to the synthetic models is useful to test the scale where the observed \textit{SF}$_{B_{\parallel}}$ flattens.

We calculated the \textit{SF}$_{B_{\parallel}}$ of the simulated fields within the same elliptical area as the observations by focusing on the data inside the same box region shown in Fig.~\ref{fig:regions}. More specifically, the simulations were conducted in a cubic domain that encloses the ellipsoid with a numerical resolution of 256$^{3}$ ($\delta x \simeq 0.2$~pc), which gives the grid  the same size as that of the observation: (major axis in pc)/(major axis in pixels) $=$ 19/104 $\sim$ 0.18 pc. We note that we show the angular values in the
final results.

Initially, we built the synthetic models with a turbulent magnetic field represented by a 3D \textit{PS} according to two power laws as
\begin{equation} \label{eq:ps}
PS_{\rm 3D}(k)\propto
    \begin{cases}
        k^{\alpha-2} & (|k| \leq k_{\rm inner}) \\
        k^{\beta-2} & (|k| > k_{\rm inner})
    \end{cases},
\end{equation}
where the wave number of the turnover $k_{inner}$ is associated with the outer scale $L_{outer}$ as \(L_{\rm outer} \equiv 2\pi/k_{\rm inner}\). We note that a LoS averaging for $n_{e}$ and $B_{\parallel}$ can also mimic a 3D spectrum \citep{Sridhar1994}; in other words,  for 3D Kolmogorov turbulence the power-law scaling slope is \(-11/3\) as \(PS_{3D}(\textbf{k}) d\textbf{k} \propto k^{-11/3}\). The full 3D magnetic field may indeed  have a Kolmogorov spectrum, but we only probe $B_{\parallel}$ here. However, according to \citet{Chepurnov1998}, if the original magnetic field has a 3D Kolmogorov signature, the parallel component may be expected to show a 3D Kolmogorov-like spectrum as well. Integrating this quantity over the LoS would then result in a 2D Kolmogorov spectrum in \textit{PS} with a power-law slope of \(-8/3\), and a 2D \textit{SF} power-law slope of \(2/3\).

Another significant aspect of turbulent $B_{\parallel}$ in this study to be noted is that it is derived from the $RM\textsubscript{Sh2--27}$ map (Fig.~\ref{fig:radio}c), and it is a LoS integration of 3D $B_{\parallel}$ weighted by $n_{e}$ and path length. Therefore, Eq.~\ref{eq:B} only holds if $n_e$ and $B_{\parallel}$ are not correlated along any LoS. In the case of a correlation (either positive or negative), this equation will generate overestimated or underestimated $B_{\parallel}$ (see \citealt{Beck2003}). Consequently, if the  \textit{SF} of $n_e$ has a Kolmogorov scaling, we can assume that Kolmogorov turbulence in $B_{\parallel}$ will result in Kolmogorov turbulence in the  \textit{SF} of \textit{RM}.

The initial $n_{e}$ is set to be constant (7.3~cm$^{-3}$) across the H\,{\sc ii} region ($n_{e}=0$ outside the ellipse). We note that we also tested a turbulent $n_{e}$ under the same conditions; however, since it gave the same \textit{SF}$_{B_{\parallel}}$ slopes, we continue our analysis with the constant $n_{e}$.

The simulated turbulent $B_{\parallel}$ follows a power law with random phases, according to Eq.~\ref{eq:ps}. The root mean square of $B_\parallel$ is set to be 6.0~$\mu$G. The value of  $L_{\rm outer}$ is set to be between 2 pc and 10~pc with a step size of 1 pc. We also computed a separate $L_{\rm outer}$ (20~pc) to test the effect of a larger scale than the computational domain on the flattening. Therefore, the Fourier components of $B_\parallel$ are drawn from a 3D Gaussian random field in Fourier space with different outer scales. The simulated $B_{\parallel}$ follows a \textit{PS}, where $\alpha$ is fixed as 2 (for \(k \leq k_{inner}\)) and \(-9/3\leq\beta\leq-5/3\), where \(\beta=-5/3\) corresponds to the Kolmogorov slope (for \(k>k_{inner}\)). By following the same procedure as for the observations,  we determined \textit{EM} and \textit{RM}, and subsequently used Eq.~\ref{eq:B} to calculate the synthetic $B_\parallel$.

After setting up all the parameters mentioned above, we ran different tests to determine the outer scale, and  to test the reliability of the simulations. Their key steps are as follows: i)  testing the effect of various outer scales on the flattening of \textit{SF}$_{B_{\parallel}}$ on large scales; ii) taking into consideration the flattening of \textit{SF}$_{B_{\parallel}}$ on small scales; iii)  addressing the effect of the different realizations of the turbulence; iv)  testing the effect of the different input slopes on the flattening. The results are the described below.

\begin{figure}
  \resizebox{\hsize}{!}
  {\includegraphics[width=9cm,height=7.4cm, center]{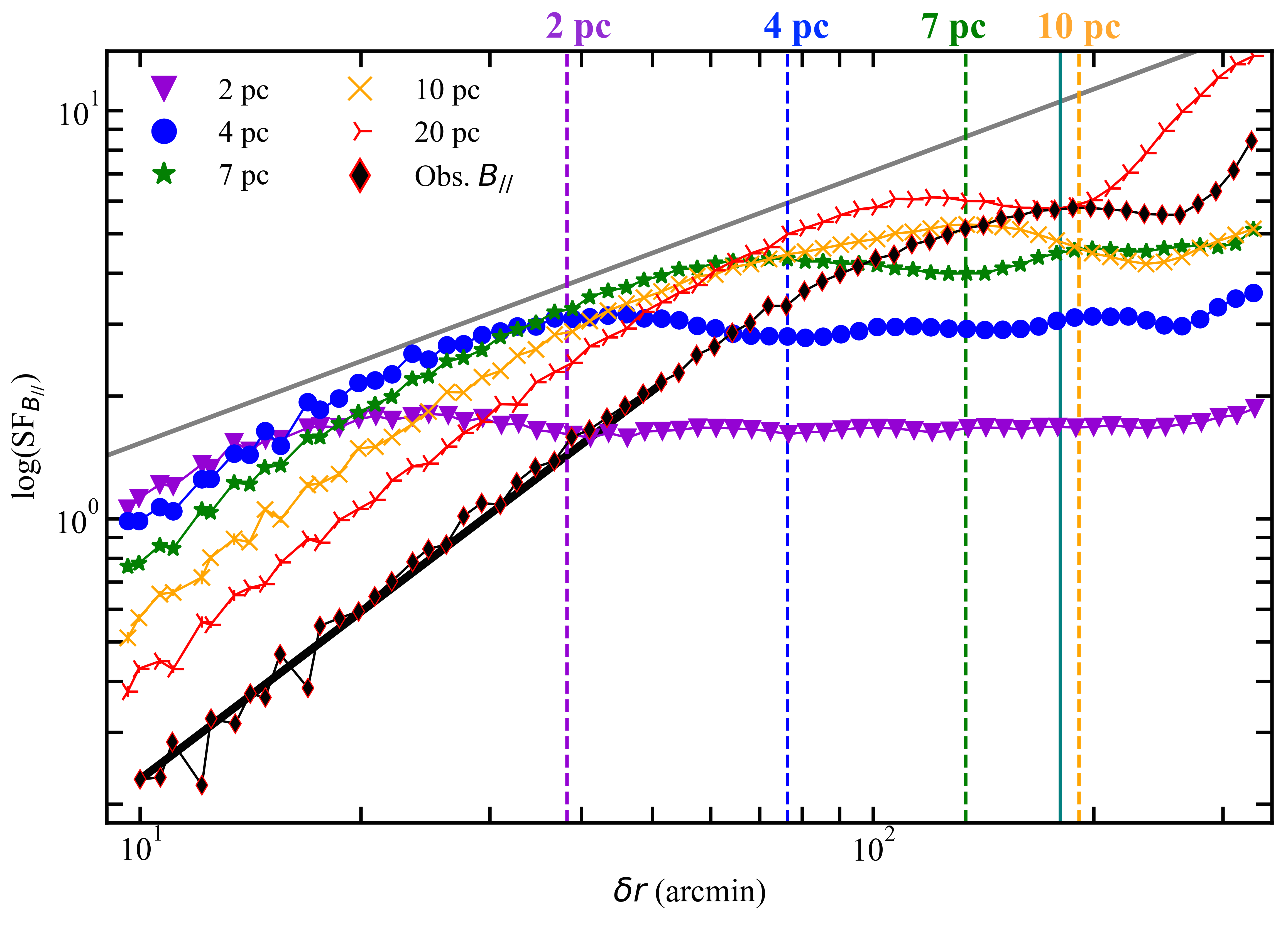}}
  \caption{Structure functions  of the observed (black) and the simulated $B_{\parallel}$ with different outer scales. The outer scales (top) are related to each \textit{SF}, coded in the same colour. The teal line represents half of the box size. The black line shows the power-law slope of 1.4. The grey line shows the \(2/3\) Kolmogorov scaling.}
  \label{fig:SFs}
 \end{figure}

In step i) we computed the simulated \textit{SF}$_{B_{\parallel}}$ with different outer scales to test the starting point of the flattening under different scenarios. In addition to the observed \textit{SF}$_{B_{\parallel}}$, Fig.~\ref{fig:SFs} shows the simulated \textit{SF}$_{B_{\parallel}}$ computed with different outer scales up to 20~pc, an input ${B_{\parallel}}$ ($-$4.5 $\mu$G),  and the  Kolmogorov slope (\(\beta=-5/3\)). The amplitudes of the simulated \textit{SF}s were set arbitrarily such that the observed \textit{SF}$_{B_{\parallel}}$ and simulated \textit{SF}$_{B_{\parallel}}$ are distinguishable in the presentation.

In step ii) the turnovers on scales  smaller  than 10~pc in the \textit{SF}$_{B_{\parallel}}$ slopes (Fig.~\ref{fig:SFs}) start flattening on scales smaller than the input outer scales. In addition, turnovers at 10~pc and 20~pc are not distinguishable  since the computational domain is limited to the size of the box.

\begin{figure}
  \resizebox{\hsize}{!}
  {\includegraphics[width=9cm,height=7.cm, center]{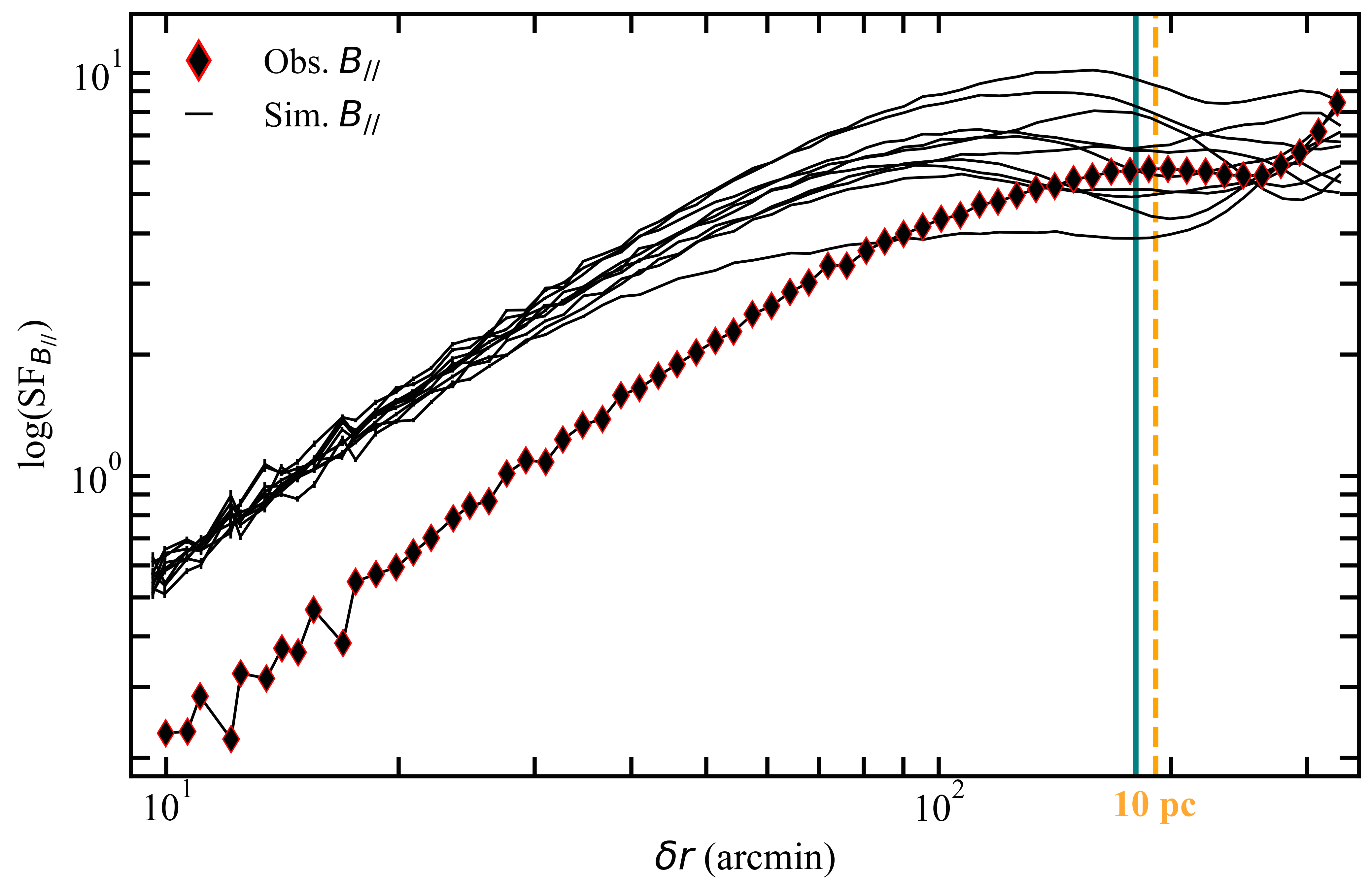}}
  \caption{Structure functions of the observed and simulated $B_{\parallel}$. Each simulated \textit{SF}$_{B_{\parallel}}$ is calculated from the different realizations of the turbulence with an outer scale of 10~pc with \(\beta=-5/3\) and \(\alpha=2\). The teal line represents half of the box size.}
  \label{fig:SFs_real}
 \end{figure}

In step iii) we test whether the different behaviours of the flattening on the large angular scales in Fig.~\ref{fig:SFs} is due to different realizations of the turbulence. We made ten representative realizations using 10~pc as the outer scale, and with an input magnetic field ($-$4.5 $\mu$G) and Kolmogorov slope (\(\beta=-5/3\)) as seen in Fig.~\ref{fig:SFs_real}. The results show that the slopes are consistent with each other on the small scales, but above the angular scale of $\sim$50$\arcmin$ the structure functions start to flatten in some realizations. Moreover, the structures on large scales ($>$50$\arcmin$) are different from each other, confirming our earlier point that they are due to inadequate sampling of the largest scales.

\begin{figure}
  \resizebox{\hsize}{!}
  {\includegraphics[width=9cm,height=7.4cm, center]{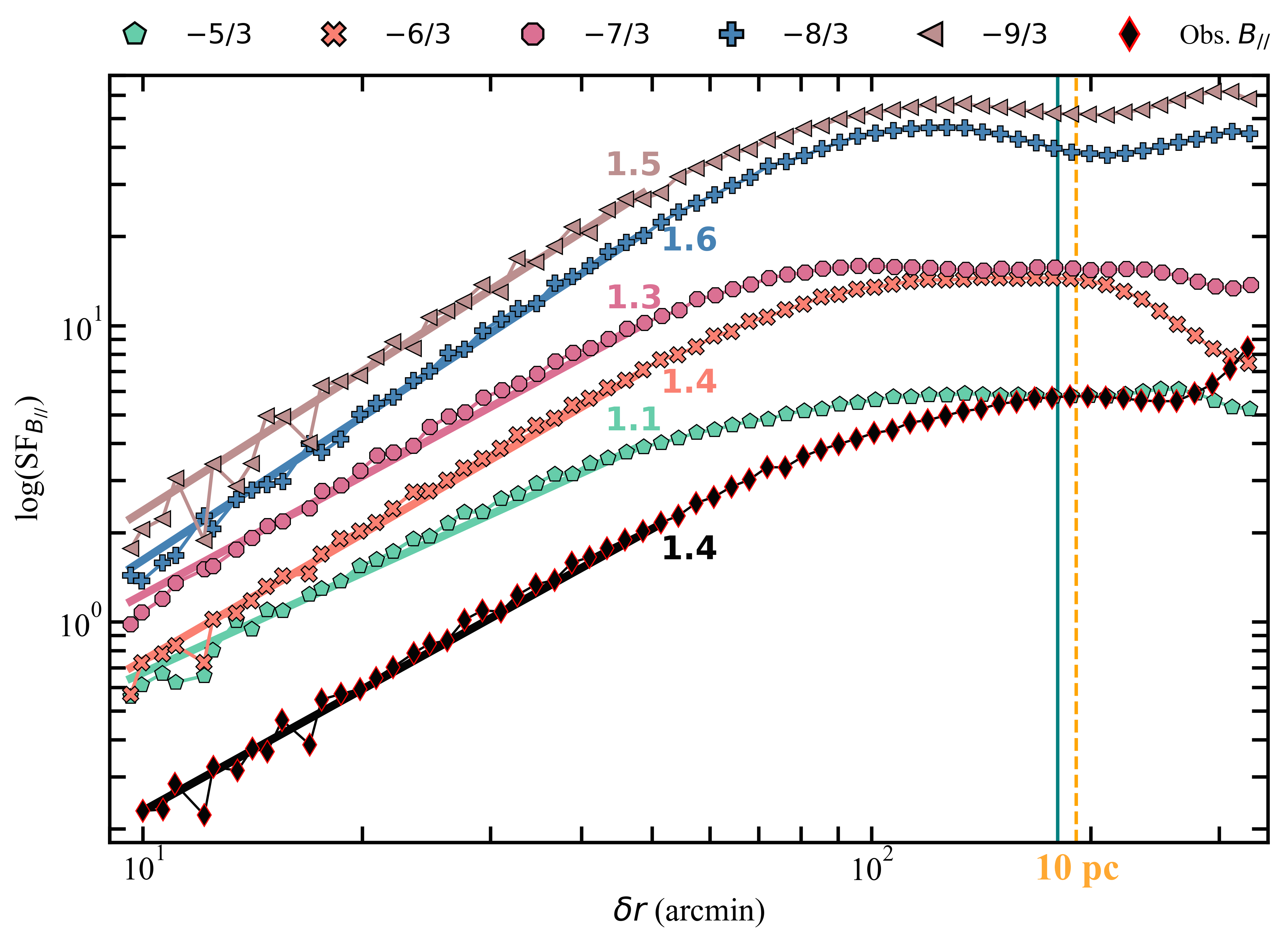}}
  \caption{Structure functions of the observed (black) and simulated $B_{\parallel}$. Each simulated \textit{SF}$_{B_{\parallel}}$ is calculated with a different input slope ($-9/3\leq\beta\leq-5/3$) with an outer scale of 10~pc, and with \(\alpha=2\). The input 1D \textit{PS} slopes of the input $B_{\parallel}$ are shown in the legend. The power-law slopes of the resulting 2D \textit{SF}s are shown next to the related data. The amplitude of each simulated \textit{SF}$_{B_{\parallel}}$ is sorted arbitrarily for a clear presentation. The teal line represents half of the box size.}
  \label{fig:SFs_steep}
 \end{figure}

In step iv), as seen in Fig.~\ref{fig:SFs}, the power-law indices of the observed and simulated \textit{SF}$_{B_{\parallel}}$ are slightly steeper than the Kolmogorov slope (\(2/3\)), up to a scale of $\sim$50$\arcmin$. A 1D \textit{PS} Kolmogorov slope of \(-5/3\) in the input magnetic field spectrum, therefore, results in a steeper output slope of \textit{SF}$_{B_{\parallel}}$. Figure~\ref{fig:SFs_steep} shows the simulations with various input magnetic field slopes ($-9/3\leq\beta\leq-5/3$), which points out that the output slopes of the simulated \textit{SF}s  slightly increase with the increasing input magnetic field slopes.

\section{Discussion}\label{sec:discussion}
In this section we interpret the results of the measured magnetic field fluctuations. In Sect.~\ref{sec:Bne_discussion} we discuss the apparent fluctuations in the magnetic field. In Sect.~\ref{sec:SF_disc} the \textit{SF}s of the magnetic field are discussed.

\subsection{$n_{e}$ and $B_{\parallel}$}\label{sec:Bne_discussion}

Several parameters affect the estimation of $n_{e}$ and $B_{\parallel}$. We  made an assumption of the LoS path length through Sh 2$-$27, and the filling factor is very uncertain. The initial assumption on the LoS path length is that it is equal to the major axis of the ellipse instead of the minor axis. Then, $\overline{n}_{e}$ and $\overline{B}_{\parallel}$ are estimated as $8.2\pm0.1$~cm$^{-3}$ and $-4.8\pm0.1~\mu$G, respectively. This means that selecting a different path length can affect the results as \(B_{\parallel} = -9.6 (\frac{38~pc}{s})^{1/2}\) $\mu$G. Similarly, to see the changes when adopting  different $f$ values, variations in both $n_{e}$ and $B_{\parallel}$ can be taken into account in proportion to $f^{-1/2}$ (see Eq.~\ref{eq:ne} and \ref{eq:B}), given that  $f$ may not be uniform across the region.

As seen in Fig.~\ref{fig:mf} in the blob-shaped extension of L204, $B_{\parallel}$ values are the lowest and $\sigma_{B_{\parallel}}$ values are the highest when $E(B-V)$ is mostly high (see Fig.~\ref{fig:emission}b). This relationship may be explained by the high values of $\tau$ leading to high $\sigma_{n_{e}}$. Additionally, it can be clearly seen that $\sigma_{B_{\parallel}}$ in the low polarized intensity regions (Fig.~\ref{fig:radio}b) is higher than in polarized regions. Hence, by evaluating Eq.~\ref{eq:ap_sigmaB2}, the analogy between $\sigma_{B_{\parallel}}$ and the depolarization canals can be shown, which points out that weak polarization signals cause the high uncertainties of $B_{\parallel}$ (see Fig.~\ref{fig:mf}b).

In Fig.~\ref{fig:mf}, multiscale structures in $B_{\parallel}$ are visible, possibly indicative of a turbulent magnetic field. Another filament-shaped structure is noticeably oriented from north-west to south-east, which is roughly parallel to the foreground polarized filament that is mentioned in Sect.~\ref{sec:analysis_interp}. We only see this filament in the $B_{\parallel}$-map. Even though it is not visible in the \textit{EM} or $RM\textsubscript{Sh2--27}$ maps, it has to be present in one or both of these, as $B_{\parallel}$ is created by these maps. As its orientation is similar to the orientation of the foreground polarized filament mentioned in Sect.~\ref{sec:pol}, it is likely that this filament is aligned with some type of large-scale magnetic field (see \citealt{Vidal2015}) in and around the H\,{\sc ii} region. The filament might be connected to the H\,{\sc ii} region or might be located in the foreground. If the latter were the case, the filament would be located in a lower-density environment, which would indicate an unrealistically high magnetic field strength to attain the observed \textit{RM} (Eq.~\ref{eq:B}). Therefore, we conclude that the filament of enhanced magnetic field strength is associated with the H\,{\sc ii} region, and pointed in a direction of a large-scale magnetic field piercing the H\,{\sc ii} region.

\subsubsection{Comparison to previous studies}
Our result of $\overline{n}_{e}$ broadly supports the previous research in Sh 2$-$27. \citet{Reynolds1982} estimated $n_{e}$ in Sh 2$-$27 as $\sim$3.8~cm$^{-3}$ using the lines of H$\alpha$ and [N\,{\sc ii}] $\lambda$6584, which is higher than our average electron density along the LoS, $\sim$1.5~cm$^{-3}$ (<$n$> $=$ $fn_{e}$). Moreover, \citet{Wood2005} found $n_{e}\sim2$~cm$^{-3}$ for Sh 2$-$27, and our result of <$n$> is compatible with their result. Additionally, 7.3$\pm$0.1 cm$^{-3}$ is consistent with the value derived by \citet{Harvey-Smith2011} within 1$\sigma$ ($n_{e}=10.6\pm2.8$~cm$^{-3}$). The slight difference is likely to be related to the estimated path length that depends on the assumed geometry of the H\,{\sc ii} region, the volume filling factor, and the  method used to  account for the contribution of the dust reddening. They assumed that Sh 2$-$27 is spherical and that the volume filling factor is 0.1, and they  employed a mean dust reddening value (0.47) from \citet{Schlegel1998}.

\begin{table*}
\caption{Values of  $|B_{\parallel}|$   estimated in other H\,{\sc ii} regions by using Faraday rotation of the polarized radio synchrotron emission observations.} 
\label{tab:B}      
\centering                                      
\begin{tabular}{lll}
\hline \hline 
\noalign{\smallskip}
H\,{\sc ii} regions                                    & $|B_{\parallel}|$ ($\mu$G)  & Reference              \\
\hline
W4                                                     & $<$20                & 1               \\
Sh 2$-$27, Sh 2$-$264, Sivan 3, Sh 2$-$171, Sh 2$-$220 & 2$-$6                & 2       \\
S117, S119, S232, S264                             & 1$-$20               & 3, 4 \\
Sh$-$205                                               & $\sim$5.7            & 5             \\
NGC 6334A                                              & $\sim$36             & 6 \\
G124.9$+$0.1, G125.6$-$1.8                             & $\sim$3.9, $\sim$6.4 & 7                \\
\hline
\end{tabular}
    \tablebib{(1) \citet{Gray1999}; (2) \citet{Harvey-Smith2011}; (3) \citet{Heiles1980}; (4) \citet{Heiles1981}; (5) \citet{Mitra2003}; (6) \citet{Rodriguez2012}; (7) \citet{Sun2007}}
\end{table*}

Our resulting $|B_{\parallel}|$ lies well within the ranges of previous studies (see Table~\ref{tab:B}) that used Faraday rotation of the polarized radio synchrotron emission observations in some H\,{\sc ii} regions. Using Faraday rotation of the extragalactic background sources, \citet{Heiles1980} and \citet{Heiles1981} estimated |$B_{\parallel}$| from 1 to 20 $\mu$G in the Galactic H\,{\sc ii} regions S117, S119, S232, and S264. \citet{Mitra2003} found |$B_{\parallel}$| of Sh 2$-$205 as $\sim$5.7~$\mu$G, using \textit{RM} and dispersion measure of the pulsars PSR J2337$+$6151 and PSR J0357$+$5236. The study of the W3/W4/W5/HB3 complex by \citet{Gray1999} specified an average upper limit for the H\,{\sc ii} region W4 as $\sim$20 $\mu$G. \citet{Sun2007} found |$B_{\parallel}$| $\sim$3.9~$\mu$G and $\sim$6.4~$\mu$G for the H\,{\sc ii} regions G124.9$+$0.1 and G125.6$-$1.8, respectively. \citet{Harvey-Smith2011} used \textit{RM}s from the \citet{Taylor2009} catalogue, and obtained |$B_{\parallel}$| from 2 to 6 $\mu$G in the five Galactic H\,{\sc ii} regions (Sh 2$-$27, Sh 2$-$264, Sivan 3, Sh 2$-$171, and Sh 2$-$220). However, H\,{\sc ii} regions of unusually high $n_{e}$ ($\sim$350~cm$^{-3}$) may have higher magnetic field, for example  $B_{\parallel}\sim$36~$\mu$G, as observed in NGC 6334A \citep{Rodriguez2012}. Another point worth noting is that the \citet{Taylor2009} catalogue is missing some sources of high \textit{RM}s in this region, which may cause biased results, and the real \textit{RM} for Sh 2$-$27 that is calculated from the extragalactic sources might be higher than these results (see \citealt{Stil2007}).

The value of $|B_{\parallel}|$ found in Sh 2$-$27 is in agreement with the diffuse ISM magnetic field (see \citealt{Crutcher2007}), which may be detected by the high $n_e$ in the H\,{\sc ii} region, as also previously pointed out by \citet{Harvey-Smith2011}.

\subsection{Structure functions}\label{sec:SF_disc}
In this section we   discuss the outer scales of fluctuations obtained and  compare them to synthetic models (Sect.~\ref{sec:derturb}), and we compare these results to previous studies (Sect.~\ref{sec:turbcomp}).

\subsubsection{Outer scales and turbulent power-law slope}\label{sec:derturb}
In the observed \textit{SF}$_{B_{\parallel}}$ the flattening occurs almost at the maximum scale of the data ($\sim$180$\arcmin$), which corresponds to a scale of $\sim$10 pc (assuming a distance to the H\,{\sc ii} region of 180 pc). The question of whether this turnover scale corresponds to the outer scale of turbulence can be answered with the help of the results of the numerical simulations in Sect.~\ref{sec:sim}. The turbulent (input) outer scales smaller than 10~pc can be seen to have smaller (output) turnover scales. In addition, no clear distinction can be made between the turnover scales for the turbulent outer scales of 10~pc and higher; see result ii) in Sect.~\ref{sec:sim}. This means that we can only give a lower limit of $\sim$10~pc to the observed maximum scale of fluctuations. If the turbulence inside the H\,{\sc ii} region is a continuation of general interstellar turbulence in the ionized environment around a massive star, the outer scale of the fluctuations may be even larger than the H\,{\sc ii} region itself. 

Result iii) in Sect.~\ref{sec:sim} shows that at scales larger than $\sim$50$\arcmin$, the effect of different realizations of the turbulence becomes noticeable, indicating that to determine  the \textit{SF}$_{B_{\parallel}}$ slopes, we should not take scales $>$50$\arcmin$ into account. 

Result iv) shows that an input 3D Kolmogorov slope \((-11/3)\) of the magnetic field in the simulations gives an output slope of 1.4, which is slightly steeper than the 2D Kolmogorov value \((>2/3)\). Figure~\ref{fig:SFs_steep} shows that  for steeper input magnetic field spectra the output \textit{SF}$_{B_{\parallel}}$ slopes also return slightly steeper values. As we stated in Sect.~\ref{sec:sim}, the input of the constant $n_e$ and turbulent $n_e$ in the simulations did not change the resulting \textit{SF}$_{B_{\parallel}}$ slope, thus the density weighting of Eq.~\ref{eq:B} is not likely to be responsible for not retrieving the \(2/3\) slope. Therefore, the reason is likely to be related to the shape of the object including an additional source of structure from the varying path length through the object, which would increase at larger scales. Nevertheless, the \textit{SF}$_{B_{\parallel}}$ slope calculated from the observations is consistent with the \textit{SF}$_{B_{\parallel}}$ slope of the simulations for a Kolmogorov-like input spectrum of the magnetic field. Therefore, we conclude that our observations are consistent with a turbulent magnetic field with a Kolmogorov slope inside Sh 2$-$27.

We note that even if the \textit{SF} slopes are consistent with the Kolmogorov turbulence, complexities still remain in translating these slopes to physical quantities such as the magnetic field. In our calculations, the following approximations induce uncertainties: (a) that the integration over the path length is approximated by $fs$; (b) that a uniform $T_{e}$ and $f$ across the region are adopted; (c) that spatial irregularities in the observed quantities of \textit{EM} and $RM\textsubscript{Sh2--27}$, and in the LoS dependency of ${B_{\parallel}}$ (\({B_{\parallel} \propto s^{-1/2}}\)) may occur.

These results suggest that \textit{SF}s of observed ${B_{\parallel}}$-maps that are computed via polarization observations can be used to characterize turbulence inside H\,{\sc ii} regions. However, this method must be approached with some caution as the size of the computational domain, adopted geometry, and stochasticity may play a significant role in the interpretation of slopes and outer scales.

\subsubsection{Comparison to previous studies}\label{sec:turbcomp}
As mentioned in Sect.~\ref{sec:sim}, we find a Kolmogorov-slope, and a lower limit to the outer scale of about 10~pc in Sh 2$-$27. If the observed turbulent magnetic field in Sh 2$-$27 is affiliated with the turbulence in the general warm ionized ISM, as argued before, it is useful to compare it to earlier studies of turbulence in this medium, considering turbulent slopes and outer scales. \citet{Stil2011} mapped out \textit{SF}$_{RM}$ across the northern sky and found a slope of $\sim$1.3 towards Sh 2$-$27, which is very close to our scaling.

The analysis of turbulent velocity structure in ionized gas is likely connected to its gas density and magnetic field fluctuations. Power spectra of velocity fluctuations in H\,{\sc ii} regions can show Kolmogorov-like spectra \citep{Roy1985,Miville1995}, although  not necessarily \citep{Odell1987,Medina1997,Chakraborty1999,Lagrois2009,Melnick2019}.

Previous works (e.g.  \citealt{Medina1997}) have also attempted to find an outer scale in an H\,{\sc ii} region by focusing on large scales. They found an outer scale of 10~pc in the extragalactic H\,{\sc ii} region NGC 604, which was attributed to a possible stellar formation event.

In the literature, Kolmogorov-like turbulence was found in the ISM in electron density fluctuations (e.g. \citealt{Armstrong1981,Higdon1984,Armstrong1990,Wang2005,Chepurnov2010}). \citet{Spangler1990} inspected the interstellar electron density \textit{PS} in radio scattering measurements, and found consistency with the predicted Kolmogorov slope of \(-11/3\). \citet{Minter1996} examined \textit{EM} and \textit{RM} fluctuations transitioned from 3D Kolmogorov-like turbulence to 2D Kolmorogov-like turbulence at large scales, and conjectured that this was due to a stratified environment of a massive star.

Even though the measured \textit{SF} slopes here are compatible with Kolmogorov-like turbulence, the turbulence in H\,{\sc ii} regions is compressible \citep{Miville1995}, which indicates that in reality the turbulence is more complex than the Kolmogorov theory.

Thus far, the previous studies on H\,{\sc ii} regions (e.g. \citealt{Miville1995,Arthur2016}) typically studied turbulent fluctuations on much smaller scales than the scales probed here. If the outer scale of the fluctuations represents the outer scale of the turbulence, it would be comparable to the size of the H\,{\sc ii} region or even exceed it. In the latter case, the turbulence we detect inside the H\,{\sc ii} region may well be a probe of the turbulence in the general ISM, which happens to be highlighted by the high electron density in the H\,{\sc ii} region \citep{Spangler2021}. 

\section{Conclusions}\label{sec:conc}
In this study, we used the Faraday rotation of the polarized radio synchrotron emission data from S-PASS at 2.3 GHz and the H$\alpha$ data from SHASSA to determine $B_{\parallel}$ in Sh 2$-$27 for each LoS. This  lead to a search for the imprint of the turbulence and its outer scale on the $B_{\parallel}$-map of the H\,{\sc ii} region  by using the second-order \textit{SF}. The following conclusions can be drawn from this study:
\begin{enumerate}
  \item[i)]By making use of three observations (linear radio polarization, H$\alpha$, and dust) we computed the maps of $n_{e}$ and $B_{\parallel}$ of Sh 2$-$27. We estimated $\overline{n}_{e}$ and $\overline{B}_{\parallel}$ in Sh 2$-$27 as $7.3\pm0.1$~cm$^{-3}$ and $-4.5\pm0.2$~$\mu$G, respectively. The $B_{\parallel}$-map for each LoS shows multiscale structures and variations ($-$18 $\mu$G < $B_{\parallel}$ < 3.5 $\mu$G) across the chosen elliptical area of the H\,{\sc ii} region. 
\item[ii)]The power-law slopes of \textit{SF}s are compatible with a Kolmogorov-like spectrum of the 3D magnetic field inside the H\,{\sc ii} region, as simulations reveal.
 \item[iii)]The observed and simulated \textit{SF}s imply that the outer scale of the turbulent fluctuations is larger than 10~pc, which is comparable to the size of the H\,{\sc ii} region. This may indicate that the turbulence probed here is the interstellar turbulence in the general ISM, cascading from the larger scales than the size of the H\,{\sc ii} region in the ambient medium, which is highlighted by Sh 2$-$27.
 
\end{enumerate}
   
By virtue of well-resolved observations, this study presents a detailed map of $B_{\parallel}$ for each LoS along with a study of the turbulent magnetic field properties in the H\,{\sc ii} region Sh 2$-$27. Further research in other H\,{\sc ii} regions would be an indispensable next step in understanding the turbulence of the magnetic field in these objects, and the ISM. 

\begin{acknowledgements}
We thank the referee for a constructive report. NCR thanks Alec J. M. Thomson for help in constructing the power spectra, Luke Pratley and Anna Ordog for discussions, and Karel D. Temmink for suggestions. This work is part of the joint NWO-CAS research programme in the field of radio astronomy with project number 629.001.022, which is (partly) financed by the Dutch Research Council (NWO). MH acknowledges funding from the European Research Council (ERC) under the European Union's Horizon 2020 research and innovation programme (grant agreement No 772663). JMS acknowledges the support of the Natural Sciences and Engineering Research Council of Canada (NSERC), 2019-04848. The Dunlap Institute is funded through an endowment established by the David Dunlap family and the University of Toronto. B.M.G. acknowledges the support of the Natural Sciences and Engineering Research Council of Canada (NSERC) through grant RGPIN-2015-05948, and of the Canada Research Chairs program. XS is supported by the National Natural Science Foundation of China (Grant No. 11763008). JH is supported by the National Natural Science Foundation of China: 11988101. XS, JH and XG are funded by the CAS-NWO cooperation programme (Grant No. GJHZ1865).

This work has made use of S-band polarization All-Sky Survey (S-PASS) data. We acknowledge the Southern H-Alpha Sky Survey Atlas (SHASSA), which is supported by the National Science Foundation \citep{Gaustad2001}.

This work made use of Astropy\footnote{\url{http://www.astropy.org}} a community-developed core Python package for Astronomy \citep{Astrocol2018}, NumPy\footnote{\url{https://numpy.org/}} \citep{Oliphant2006,vanderWalt2011}, and Matplotlib\footnote{\url{http://www.matplotlib.org/}} python module \citep{Hunter2007}. We have made use of the ``inferno'' and ``bone'' colour-maps, and colour-blind-friendly figures.
\end{acknowledgements}

\begin{center}
ORCID IDs
\end{center}

N. C. Raycheva\hspace{1mm}\includegraphics[width=7pt]{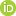}\hspace{1mm}\href{https://orcid.org/0000-0003-2859-0527}{\color{blue}https://orcid.org/0000-0003-2859-0527}

M. Haverkorn\hspace{1mm}\includegraphics[width=7pt]{images/orcid_16x16.png}\hspace{1mm}\href{https://orcid.org/0000-0002-5288-312X}{\color{blue}https://orcid.org/0000-0002-5288-312X}

J. M. Stil\hspace{1mm}\includegraphics[width=7pt]{images/orcid_16x16.png}\hspace{1mm}\href{https://orcid.org/0000-0003-2623-2064}{\color{blue}https://orcid.org/0000-0003-2623-2064}

B. M. Gaensler\hspace{1mm}\includegraphics[width=7pt]{images/orcid_16x16.png}\hspace{1mm}\href{https://orcid.org/0000-0002-3382-9558}{\color{blue}https://orcid.org/0000-0002-3382-9558}

J. L. Han\hspace{1mm}\includegraphics[width=7pt]{images/orcid_16x16.png}\hspace{1mm}\href{https://orcid.org/0000-0002-9274-3092}{\color{blue}https://orcid.org/0000-0002-9274-3092}

E. Carretti\hspace{1mm}\includegraphics[width=7pt]{images/orcid_16x16.png}\hspace{1mm}\href{https://orcid.org/0000-0002-3973-8403}{\color{blue}https://orcid.org/0000-0002-3973-8403}

%
%

\begin{appendix}
\section{Propagation of errors in estimations} \label{app:app}

We included the errors for each data point in our estimations by adapting the propagation of error calculations \citep{Squires2001,Brentjens2005,Hughes2010}. 

We derive the uncertainty of polarization angles following \citet{Brentjens2005} as
\begin{equation}
\label{eq:ap_sigmapsi}
\sigma_{\chi}^2 \ = \ \left(\frac{\partial\chi}{\partial
Q}\right)^2\sigma_\mathrm{Q}^2 + \left(\frac{\partial \chi}{\partial
U}\right)^2\sigma_\mathrm{U}^2,
\end{equation}
where \(\sigma _{Q}=\sigma _{U}=\sigma _{QU}\), which  leads to \(\sigma_{\chi} \sim \sigma _{RM\textsubscript{Sh2--27}}\) by
\begin{equation}
\label{eq:ap_sigmarm}
\sigma_{RM}^2 \ = \ \left(\frac{\partial RM\textsubscript{Sh2--27}}{\partial \chi}\right)^2 \sigma^{2}_{\chi}.
\end{equation}
The uncertainty of \textit{EM} is estimated as
\begin{equation}
\sigma_{EM}^2 \ = \ \left(\frac{\partial EM}{\partial I_{H\alpha}}\right)^2 \sigma^{2}_{I_{H\alpha}} + \left(\frac{\partial EM}{\partial \tau}\right)^2 \sigma^{2}_{\tau},
\end{equation}
where \(\sigma_{I_{H\alpha}}=0.6\) R for each data point (after convolution to S-PASS), and $\sigma_{\tau}$ can be derived from the dust map by using the Python {\sc statistics} module. Within the same adaption the uncertainty of $n_{e}$ is
\begin{equation}
\sigma_{n_{e}}^2 \ = \ \left(\frac{\partial n_{e}}{\partial EM}\right)^2 \sigma^{2}_{EM}.
\end{equation}
Finally, the uncertainty of $B_{\parallel}$ is
\begin{equation}
\label{eq:ap_sigmaB}
\sigma_{B_{\parallel}}^2 \ = \ \left(\frac{\partial B_{\parallel}}{\partial EM}\right)^2 \sigma^{2}_{EM} + \left(\frac{\partial B_{\parallel}}{\partial RM\textsubscript{Sh2--27}}\right)^2 \sigma^{2}_{RM\textsubscript{Sh2--27}},
\end{equation}
which can also be rewritten as
\begin{equation}
\label{eq:ap_sigmaB2} 
\frac{\sigma_{B}^2}{B_{\parallel}^2}  = \frac{\sigma_{RM\textsubscript{Sh2--27}}^{2}}{RM\textsubscript{Sh2--27}^{2}} + \frac{\sigma_{EM}^{2}}{4EM^{2}},
\end{equation}
so that if we consider \(\sigma _{\chi}^{2} = \sigma _{P}^{2}/(4P^{2})\) from   Eq. A.12 in \citealt{Brentjens2005}, the relationship between the uncertainties of the magnetic field and the depolarization canals, in our study, can be found from \(\sigma _{RM}^{2} \sim \sigma _{P}^{2}/(4P^{2}\lambda^{4})\), where
\begin{equation}
\sigma_{P}^2 = \left(\frac{\partial P}{\partial Q}\right)^2 \sigma_{Q}^{2} + \left(\frac{\partial P}{\partial U}\right)^2\sigma_{U}^{2}.
\end{equation}
\end{appendix}

\end{document}